\documentclass[eqsecnum,nofootinbib,12pt]{revtex4-1}
\pdfoutput=1

\makeatletter
\renewcommand{\p@subsection}{}
\renewcommand{\p@subsubsection}{}
\makeatother

\usepackage{graphicx}
\usepackage{amsmath}
\usepackage{slashed}
\usepackage{amssymb}
\usepackage[mathscr]{euscript}
\usepackage{enumerate}
\usepackage{enumitem}
\usepackage{hyperref}

\hypersetup{
	colorlinks=true,
	linktoc=page,
	linkcolor=blue,
	citecolor=blue,
	urlcolor=blue
	}

\newlist{Llist}{enumerate}{3}
\setlist[Llist, 1]{
   label*=(L\arabic*),
   }

\newlist{Alist}{enumerate}{3}
\setlist[Alist, 1]{
   label*=(A\arabic*),
   }

\newlist{Blist}{enumerate}{3}
\setlist[Blist, 1]{
   label*=(B\arabic*),
   }

\newlist{Clist}{enumerate}{3}
\setlist[Clist, 1]{
   label*=(C\arabic*),
   }

\newlist{Dlist}{enumerate}{3}
\setlist[Dlist, 1]{
   label*=(D\arabic*),
   }

\DeclareMathOperator{\Tr}{Tr}
\newcommand{\Sc}[0]{ {\mathscr S} }
\newcommand{\ScOP}[0]{ {\mathscr S}^{(1)} }
\newcommand{\ScChi}[0]{ \Sc^{(\hat\chi)} }

\newcommand{\Ssc}[0]{ {\mathfrak S}^{(1)} }
\newcommand{\ph}[0]{ \phi }
\newcommand{\Ph}[0]{ \Phi }
\newcommand{\bdiamond}[0]{ \blacklozenge }

\begin{document}

\title{Coarse-grained entropy and causal holographic information in AdS/CFT}

\author{William R. Kelly}
\email{wkelly@physics.ucsb.edu}
\author{Aron C. Wall}
\email{aronwall@physics.ucsb.edu}
\affiliation{University of California at Santa Barbara, Santa Barbara, CA 93106, USA}
\date{\today}

\begin{abstract}
We propose bulk duals for certain coarse-grained entropies of boundary regions.
The `one-point entropy' is defined in the conformal field theory by maximizing the entropy in a domain of dependence while fixing the one-point functions.  We conjecture that this is dual to the area of the edge of the region causally accessible to the domain of dependence (i.e. the `causal holographic information' of Hubeny and Rangamani).
The `future one-point entropy' is defined by generalizing this conjecture to future domains of dependence and their corresponding bulk regions.  We show that the future one-point entropy obeys a nontrivial second law.   If our conjecture is true, this answers the question ``What is the field theory dual of Hawking's area theorem?''

\end{abstract}

\maketitle

\tableofcontents

\newpage

\section{Introduction}

The AdS/CFT correspondence predicts that the effective degrees of freedom of certain conformal field theories (CFT's) in the large $N$ limit are the same as the degrees of freedom of classical supergravity~\cite{Maldacena:1997re,*Gubser:1998bc,*Witten:1998qj}.  Despite many nontrivial tests of the correspondence, the precise way in which local interactions emerge in the large $N$ limit of strongly coupled CFT's is not fully understood.  What is known is that locality in the holographic dimension is intimately connected with the locality of the renormalization group (RG) flow in the CFT~\cite{Swingle:2009bg,Heemskerk:2010hk,Faulkner:2010jy,Lee:2013dln}.  From a Wilsonian point of view, this suggests that the emergence of locality in the bulk theory is related to some kind of coarse graining in the CFT.

One technical difficulty with making this idea precise is choosing an appropriate regulator to cut off the high energy modes.  This problem is particularly difficult in the physically correct Lorentz signature.  There the elimination of highly boosted modes normally requires sacrificing either Lorentz invariance (e.g.  with a hard energy cutoff), or else positivity of the inner product (e.g. Pauli-Villars~\cite{Pauli:1949zm}).  On the other hand, the bulk theory is Lorentz-invariant, and presumably has positive probabilities.    Thus, although there is detailed \emph{qualitative} agreement between the dependence of fields in the radial direction, and the RG flow of the field theory, a comprehensive framework relating the two is lacking.

Similar problems arise in the context of thermodynamics.  In order to obtain a nontrivial second law of thermodynamics, one needs to define a coarse-grained entropy.  As with the renormalization group flow, there are multiple possible coarse graining procedures.  Which one you choose affects the exact results for quantities like the entropy, introducing an element of subjectivity.  One hopes that in the thermodynamic limit, the choice does not matter at leading order.  But gauge/gravity duality suggests that (at least in the large $N$ limit) there may be a particular coarse graining procedure which has especially nice properties, due to its relation to bulk locality.

In this article we will explore the relation between coarse graining of the CFT and bulk locality.  Rather than focusing on the RG flow, we will study the localization of information in the CFT by attempting to relate coarse-grained entropies in regions of the CFT to areas of bulk surfaces.

We take inspiration from the Ryu-Takayanagi conjecture (and its later generalization by Hubeny, Rangamani, and Takayanagi) which relates the \emph{fine-grained} von Neumann entropy of a piece of the boundary to the area of minimal or extremal/maximin surfaces in the bulk known as the holographic entanglement entropy~\cite{Ryu:2006bv,Ryu:2006ef,Hubeny:2007xt,Wall:2012uf}.  This conjecture has been validated in every case in which we have control over the calculations on both sides of the duality and significant progress has been made towards a proof \cite{Fursaev:2006ih,Headrick:2010zt,Casini:2011kv,Lewkowycz:2013nqa,Hartman:2013mia,Faulkner:2013yia}.  Work has even begun on explicit constructions of the bulk geometry from the holographic entanglement entropy of arbitrary boundary regions~\cite{Hammersley:2006cp,Hammersley:2007ab,Bilson:2008ab,Nozaki:2013vta,Lashkari:2013koa,Bhattacharya:2013bna}.  
Here we will propose a similar conjecture, but using a \emph{coarse-grained} entropy of a boundary region, in place of the von Neumann entropy.

More recently Hubeny and Rangamani proposed a new quantity $\chi_{\cal A}$ which they called the ``causal holographic information"~\cite{Hubeny:2012wa,Hubeny:2013hz,Hubeny:2013gba}.  This quantity is equal to the area of a co-dimension two surface in the bulk that is defined by its casual relation to a boundary region $\cal A$.  For a host of reasons Hubeny and Rangamani conjectured that $\chi$ quantifies some aspect of the information content of the associated boundary domain of dependence.\footnote{See also~\cite{Bousso:2012sj,Czech:2012bh} for other approaches to understanding the information contained in boundary regions.}  We will present evidence that, for source-free boundary theories, $\chi$ is dual to a particular coarse-grained entropy $\ScOP$.  We will refer to $\ScOP$ as the `one-point entropy', because it depends only on the one-point functions of local operators in the domain of dependence of $\cal A$.

We also propose a second duality between a coarse graining $\Ssc$ (the `future one-point entropy') and a bulk quantity $\ph$ (the `future causal information').  These quantities are natural generalizations of $\ScOP$ and $\chi$, but have the appealing new property that they can increase during processes which involve thermalization in the CFT (corresponding to horizon formation in the bulk).  If this new conjecture is correct, the thermodynamic second law obeyed by $\Ssc$ is dual to the area theorem in general relativity~\cite{Hawking:1971tu}, as applied to causal horizons of the form $\partial J^-({\cal Z})$ where $\cal Z$ is some set of points on the boundary of AdS and $\partial J^-$ is the boundary of the causal past.\footnote{This generalizes the notion of `causal horizon' defined by Jacobson and Parentani~\cite{Jacobson:2003wv}, whose definition would require ${\cal Z}$ to be just one point.}  In this way we propose a precise connection between Hawking's area theorem and the thermalization of a quantum mechanical system.

In section~\ref{sec:review} we briefly review the definition of the causal holographic information and establish our notation.  In section~\ref{sec:CGE} we define a class of coarse-grained entropies and explore their general properties.  In section~\ref{sec:OPentropy} we define the one-point entropy $\ScOP$ and present evidence for the conjecture that $\ScOP = \chi$ (for source-free boundary theories).  We also comment on the uniqueness of our proposal and the prospects for precision tests.  In section~\ref{sec:StrongCG} we define the future causal information $\ph$ and the future one-point entropy $\Ssc$ and present evidence that they are also dual to each other (for source-free boundary theories).  Finally, in section~\ref{sec:Alt} we conclude by summarizing our results and commenting on the prospects of extending our conjectures to the semiclassical regime.

Appendix~\ref{app:chiPreserving} presents two illustrative examples of failed proposals for the dual of $\chi$, and appendix~\ref{app:BoundarySources} constructs a counterexamples to our conjecture, in the case where boundary sources are allowed.

Whenever possible we adopt the notation of \cite{Hubeny:2012wa} (see section~\ref{sec:review} for a review) with the exception that we use $D^\pm[{\cal A}], J^\pm[{\cal A}]$ to refer to the boundary future (past) domain of dependence and domain of influence and $D_\textrm{bulk}^\pm[{\cal A}], J_\textrm{bulk}^\pm[{\cal A}]$ to refer to the associated bulk regions.

\section{Causal holographic information: A brief review}\label{sec:review}

\begin{figure}
\includegraphics[width=0.3 \textwidth]{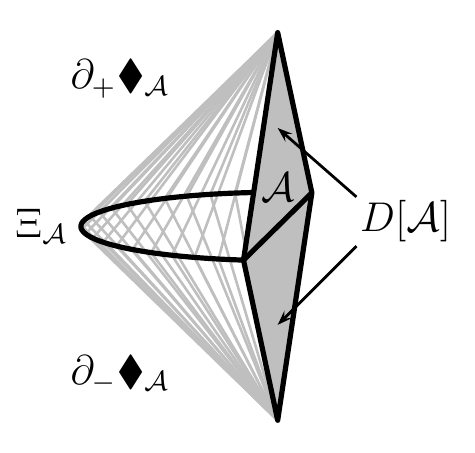} 
\caption{A sketch of the causal wedge construction of~\cite{Hubeny:2012wa}.  $D[{\cal A}]$ is the boundary domain of dependence of $\cal A$ and $\Xi_{\cal A}$ extends into the bulk (see text).}
\label{fig:causalwedge}
\end{figure}

In this section we briefly review the definition of causal holographic information $\chi$.  See \cite{Hubeny:2012wa,Hubeny:2013hz,Hubeny:2013gba} for additional details. We emphasize that for our purposes, $\chi$ is only well-defined on classical geometries (i.e. in the strict $N\rightarrow\infty$ limit).

Consider a closed spatial region ${\cal A}$ on the boundary CFT of an asymptotically AdS spacetime.\footnote{Since we are restricting to source-free boundaries, we only consider the case in which the boundary is conformally flat.  But perhaps it is possible to generalize to static boundary geometries.}  We assume that ${\cal A}$ is achronal (i.e. no timelike curves pass through it more than once), and codimension-one on the boundary.  The region ${\cal A}$ defines a causal domain of dependence $D[{\cal A}] = D^+[{\cal A}] \cup D^-[{\cal A}]$, where $D^\pm[{\cal A}]$ is defined as the collections of points $p$ for which any infinitely extended timelike curve must intersect $\cal A$ to the past (future) of~$p$~\cite{Geroch:1970uw}.

The boundary domain of dependence $D[{\cal A}]$ defines a bulk causal wedge:
\begin{align}
\bdiamond_{\cal A} = J^+_\textrm{bulk}[D[{\cal A}]] \cap J^-_\textrm{bulk}[D[{\cal A}]],
\end{align}
where $J^\pm_\textrm{bulk}[{\cal A}]$ is the future (past) of $D[{\cal A}]$ in the bulk.  In other words any point $p$ in $\bdiamond_{\cal A}$ lies on at least one causal curve that begins and ends in $D[{\cal A}]$ (see Fig.~\ref{fig:causalwedge}).

Even though the topology of $\bdiamond_{\cal A}$ may be nontrivial~\cite{Hubeny:2013gba}, 
the boundary of $\bdiamond_{\cal A}$ can be written as
\begin{align}
\partial \, \bdiamond_{\cal A}  = \partial_+ \bdiamond_{\cal A} \cup \partial_- \bdiamond_{\cal A},
\end{align}
where $\partial_\pm \bdiamond_{\cal A}$ are future (past) horizons anchored to the future (past) boundary of $D[{\cal A}]$.  These null surfaces intersect in a co-dimension two surface
\begin{align}
\Xi_{\cal A} = \partial_+ \bdiamond_{\cal A} \cap \partial_- \bdiamond_{\cal A},
\end{align}
known as the `causal information surface' from which we calculate the causal holographic information:
\begin{align} \label{eq:CHIdef}
\chi_{\cal A} = \frac{\mathrm{Area}[ \Xi_{\cal A} ]}{4 G_N},
\end{align}
where $G_N$ is Newton's constant.

Equation~\eqref{eq:CHIdef} is reminiscent of the definition of the HEE:
\begin{align} \label{eq:HEEdef}
S_{\cal A} = \frac{\mathrm{Area}[{\mathfrak E}_{\cal A}]}{4 G_N},
\end{align}
where ${\mathfrak E}_{\cal A}$ is defined as the minimum area extremal surface homologous to $\cal A$~\cite{Hubeny:2007xt} or equivalently as the maximin surface as described in~\cite{Wall:2012uf}.  We mention here, since it will come up many times in our later analysis, that it has been shown in~\cite{Hubeny:2012wa,Wall:2012uf} that 
\begin{align} \label{eq:Sgtrchi}
S_{\cal A} \le \chi_{\cal A}
\end{align}
for smooth spacetimes satisfying the null energy condition which we will assume throughout, since we are concerned with supergravity theories arising in AdS/CFT, for which the null energy condition holds classically.

Throughout this paper we will assume that the Ryu-Takayanagi conjecture is true.  More precisely we assume that the order $N^2$ contribution to the von Neumann entropy of the reduced density matrix on $\rho_{\cal A}$ is equal to $S_{\cal A}$.\footnote{Here we gloss over subtle questions involving how to define local observables in a gauge theory, and whether there are additional ``contact terms'' besides the entanglement entropy which should be included in the definition of $S_{\cal A}$~\cite{Kabat:1995eq,Larsen:1995ax,Iellici:1996jx,Zhitnitsky:2011tr,Donnelly:2012st,Solodukhin:2012jh,Eling:2013aqa,Donnelly:2013}.}  Since we will only ever be interested in the $N \rightarrow \infty$ limit (see section~\ref{sec:CorrPrinc} below) we will avoid introducing a new symbol and simply let
\begin{align} \label{eq:vonNeumann}
S_{\cal A}(\rho_{\cal A}) = - \Tr[\rho_{\cal A} \log(\rho_{\cal A})].
\end{align}

Note that the entanglement entropy is divergent, as is the area of ${\mathfrak E}_{\cal A}$.  In principle, one should figure out what is the precise numerical relationship between the two cutoffs, in order to compare the bulk and boundary quantities using the UV/IR correspondence~\cite{Susskind:1998dq}.  Since this is difficult, it is more usual to cut off both quantities independently, and then to compare only quantities which are independent of the cutoff procedure~\cite{Ryu:2006bv,Casini:2011kv}.  This includes logarithmic divergences and certain finite terms.  Note also that the divergences are state independent (at least for regular states), so universal information can also be extracted by comparing states.

Presumably, a similar procedure should be used for $\chi_{\cal A}$ and $\ScOP_{\cal A}$.  However, unlike ${\mathfrak E}_{\cal A}$, the divergences in the area of $\Xi_{\cal A}$ depend on the choice of ${\cal A}$ in a nonlocal way~\cite{Freivogel:2013zta}.  We will comment briefly in section~\ref{sec:tests} on the plausibility of $\ScOP_{\cal A}$ and $\chi_{\cal A}$ having matching divergences.  Note that because $\chi$ and $S$ differ in their divergences, inequalities such as $S_{\cal A} \le \chi_{\cal A}$ typically reduce to a statement comparing the coefficients of their leading-order divergences.\footnote{This requires that the quantities be regulated in a manner consistent with the proof; for example theorem~14 of~\cite{Wall:2012uf} compares the surfaces $\Xi$ and ${\mathfrak E}$ using the second law, so the two surfaces must be regulated in such a way that the second law can be used.}

\section{Coarse-grained entropies}\label{sec:CGE}

\subsection{Definition}

For the purposes of this paper a coarse-grained entropy is calculated by maximizing the von Neumann entropy subject to some set of constraints.  More precisely, we define a coarse-grained entropy $\Sc_{\cal A}$ associated with boundary region $\cal A$ to be (cf. \cite{GellMann:2006uj})
\begin{align} \label{eq:ScDef}
\Sc_{\cal A}(\rho_{\cal A}) =  \sup_{\tau_{\cal A} \in  T_{\cal A} }
\left[ S_{\cal A}(\tau_{\cal A}) \right]  
\end{align}
where $\rho_{\cal A}$ is the reduced density matrix associated with $\cal{A}$, $S_{\cal A}(\tau_{\cal A})$ is the von Neumann entropy of $\tau_{\cal A}$, and
$ T_{\cal A}(\rho_{\cal A})$ is the set of all density matrices $\tau_{\cal A}$ which satisfy the constraints
\begin{align} \label{eq:CGconstraint}
\Tr[{\cal O}_m \, \tau_{\cal A}] = \Tr[{\cal O}_m \rho_{\cal A}] 
\end{align}
where the $\{ {\cal O}_m \}$ are a set of operators supported in $D[{\cal A}]$.  Different coarse-grained entropies differ only in the choice of constraints.

We will call the density matrix $\sigma_{\cal A} \in  T_{\cal A}$ that maximizes the von Neumann entropy the ``coarse graining'' of $\rho_{\cal A}$, so that
\begin{align}
\Sc_{\cal A}(\rho_{\cal A}) = S_{\cal A}(\sigma_{\cal A}).
\end{align}
This coarse-grained state must be unique, since if we had two candidate states with equal entropy $\sigma_{\cal A}^{(1)}$ and $\sigma_{\cal A}^{(2)}$, then by convexity of the von Neumann entropy we could construct a higher entropy state $\sigma_{\cal A} = (\sigma_{\cal A}^{(1)} + \sigma_{\cal A}^{(2)})/2$.  According to \cite{GellMann:2006uj} the general solution to~\eqref{eq:ScDef} is (even when the ${\cal O}_m$ are not mutually commuting)
\begin{align} \label{eq:sigmaSol}
\sigma_{\cal A} = Z^{-1} \exp\left( - \sum_m \lambda_m {\cal O}_m \right),
\end{align}
where $\lambda_m$ are Lagrange multipliers determined by solving~\eqref{eq:CGconstraint} and the normalization constant $Z$ is the partition function.  In other words $\sigma_{\cal A}$ is a sort of generalized ensemble in which the $\lambda_m$ play the role of chemical potentials.

It will be useful in the following discussion to characterize coarse grainings by their relative strengths as follows.  Consider two entropies $\tilde\Sc$ and $\bar\Sc$ as defined above with different sets of constraints.  If the constraints of $\tilde\Sc$ are a proper subset of the constraints of $\bar\Sc$ (so that $\bar T \subset \tilde T$) then we say that $\tilde\Sc$ is a stronger coarse graining than $\bar\Sc$ and we use the notation $\bar\Sc \prec \tilde\Sc$.\footnote{Note that when the constraints are weaker, the coarse graining is ``stronger", in that one is forgetting more about the state.  The weakest possible coarse graining is simply the fine-grained entropy $S$, which involves constraining all information about the state.}   This implies that
\begin{align} \label{eq:compent}
\bar\Sc_{\cal A}(\rho_{\cal A}) \le \tilde\Sc_{\cal A}(\rho_{\cal A}), 
\end{align}
for all states $\rho_{\cal A}$, where equality holds if and only if $\tilde\sigma_{\cal A} \in \bar T(\rho_{\cal A})$.  Finally, if for two coarse grainings $\hat\Sc$ and $\bar\Sc$ neither set of constraints is a subset of the other, then we say that $\hat\Sc$ and $\bar\Sc$ are incomparable and we use the notation $\hat \Sc \parallel \tilde \Sc$.

For future reference we prove a mathematical result that holds for all $ \Sc$:
\begin{Llist}
\item \label{lemma}
For any positive definite, Hermitian density matrix we may, without loss of generality, write
\begin{align}
\rho_{\cal A} = Z^{-1} \exp( - \beta H). \label{thermal}
\end{align}
The operator $H$ is known as the modular Hamiltonian associated with $\rho_{\cal A}$ and is generally non-local except in a few special cases, $\beta$ is a number, and $Z =  \Tr[ \exp( - \beta H)]$.  If $H $ is one of the constraint operators associated with $\Sc$, (i.e. $ H \in \{ {\cal O}_m \}$) then
\begin{align} \label{eq:thermalSc}
\Sc_{\cal A}(\rho_{\cal A}) = S_{\cal A}(\rho_{\cal A}).
\end{align}

The proof is as follows:  The state $\rho_{\cal A}$ maximizes the entropy subject to a subset of the constraints (namely the constraint associated with $\left< H \right>$), but adding additional constraints can only lower the entropy, therefore
\begin{align}
\Sc_{\cal A}(\rho_{\cal A}) \le S_{\cal A}(\rho_{\cal A}).
\end{align}
However, $\rho_{\cal A}$ satisfies all of the constraints~\eqref{eq:CGconstraint}; therefore by virtue of the maximization condition in~\eqref{eq:ScDef} we also have
\begin{align}
\Sc_{\cal A}(\rho_{\cal A}) \ge S_{\cal A}(\rho_{\cal A}),
\end{align}
and thus we obtain \eqref{eq:thermalSc}.

\end{Llist}

\subsection{A correspondence principle} \label{sec:CorrPrinc}

Whereas the coarse-grained entropies $\Sc$ are defined for all reduced density matrices $\rho_{\cal A}$, $\chi$ is defined only on classical spacetimes.  This means that any correspondence between some $\Sc$ and $\chi$ must be restricted to the large $N$ limit of the dual field theory.  More precisely we define the correspondence limit of a coarse-grained entropy by calculating $\Sc$ at finite $N$ and retaining only the order $N^2$ term as we formally take the $N\rightarrow \infty$ limit.  We will work in the general relativity limit, in which the bulk Newton's constant $G_N$ remains finite as the string and Planck lengths vanish.  Of course, it would be of interest to extend the definition of $\chi$ into the semiclassical regime perhaps using the generalized entropy~\cite{Bekenstein:1973ur,*Hawking:1974sw} as inspiration (see~\cite{Wall:2009wm} for an extensive review) and compare subleading corrections; however we will not pursue that idea in this work except for brief comments in section~\ref{sec:Alt}.

Of course not every density matrix is dual to a classical geometry in the bulk.  We will therefore be particularly interested in density matrices which define a bulk causal wedge $\bdiamond_{\cal A}$ in the dual description.  We will call any such density matrix a ``classical state."  Note that if $\rho_{\cal A}$ is classical it is not clear that the coarse-grained state $\sigma_{\cal A}$ must also be classical.

A subtlety arises when $\cal C$ is a Cauchy surface of the boundary, i.e. when $D[{\cal C}]$ is the entire boundary.  In this case, the field theory states will experience Poincar\'e recurrences and other large fluctuations over times of order $\exp(N^2)$.  These fluctuations and recurrences allow thermal states to be reconstructed simply by waiting an extremely long time.  It is therefore appropriate that in the correspondence limit we monitor the constraints~\eqref{eq:CGconstraint} only over times that are parametrically larger then any scale in the classical spacetime, while still being parametrically smaller than $\exp(N^2)$.

\begin{figure}
\includegraphics[width=0.3\textwidth]{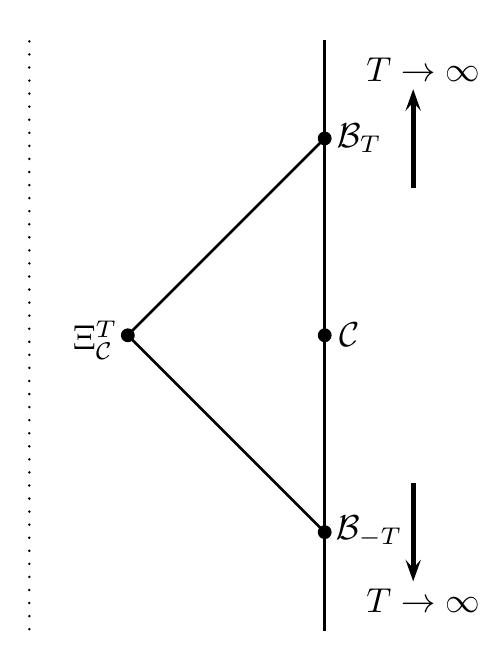} 
\caption{When $\cal C$ is a Cauchy surface $\chi_{\cal C}$ is calculated from the area of $\Xi^T_{\cal C}$.  ${\cal B}_{\pm T}$ are slices of a foliation of boundary Cauchy surfaces and $\Xi^T_{\cal C}$ is the intersection of their respective past and future horizons.  This construction addresses non-perturbative late time quantum effects such to Poincar\'e recurrences and black hole evaporation.}
\label{fig:limit}
\end{figure}

More precisely we define $\Sc_{\cal C}$ by introducing a foliation of Cauchy surfaces ${\cal B}_t$ and replacing $D[{\cal C}]$ with region bounded by ${\cal B}_{-T}$ and ${\cal B}_T$.  We then take $T \rightarrow \infty$ as $N\rightarrow \infty$ while maintaining $T \ll \exp(N^2)$.\footnote{or using the much shorter black hole evaporation time for spacetimes with sufficiently small black holes.}  On the bulk side we use the same foliation ${\cal B}_t$ of the boundary to define the family of surfaces (see Fig.~\ref{fig:limit})
\begin{align}
\Xi^T_{\cal C} = \partial_+ (J_\mathrm{bulk}^-[{\cal B}_T] ) \cap  \partial_- (J_\mathrm{bulk}^+[{\cal B}_{-T}]),
\end{align}
and we define the causal holographic information of the Cauchy surface $\cal C$ as
\begin{align} \label{eq:chiLimit}
\chi_{\cal C} = \lim_{T\rightarrow\infty} \frac{\textrm{Area}\left[\Xi^T_{\cal C}\right ]}{4 G_N}.
\end{align}

One consequence of taking the correspondence limit is that it is possible for coarse grainings which are different at finite $N$ to agree to order $N^2$ for all classical states as we take $N\rightarrow \infty$.  We will say that any two such coarse grainings are ``equivalent" and we will use the symbol $\bar\Sc \equiv \tilde\Sc$.\footnote{This fact suggests a more general class of coarse grainings.  One could replace the constraint~\eqref{eq:CGconstraint} with
\begin{align} \label{eq:approxConstraint}
\left| \Tr[{\cal O}_m \tau_{\cal A}] - \Tr[{\cal O}_m \rho_{\cal A}] \right| < c_m N^{1-k_m},
\end{align}
where $c_m,k_m$ are positive constants.  It is then possible that these generalized coarse grainings would agree with our coarse grainings $\Sc$ in the correspondence limit, but differ for finite $N$.  Coarse grainings of this type could play an important role in future investigations of the semiclassical regime.  For now, however, we will only use constraints of the form~\eqref{eq:CGconstraint} because we are uncertain how to choose $c_m$ and $k_m$.  We thank Don Marolf for pointing this out.\label{foot:approx}}  We will often only be interested in classifying coarse-grained entropies as stronger or weaker up to this equivalence relation.

\subsection{General properties } \label{sec:genProp}

We now list a few general properties that hold for all coarse-grained entropies $\Sc$.

\begin{Alist}

\item \label{genprop:diamond}
{ \bf The coarse-grained entropy of $\cal A$ depends only on the domain of dependence $D[{\cal A}]$:}  In particular, if there are two regions $\cal A$ and $\cal B$ for which $D[{\cal A}] = D[{\cal B}]$ then $\rho_{\cal A} = \rho_{\cal B}$ and $\Sc_{\cal A}(\rho_{\cal A}) = \Sc_{\cal B}(\rho_{\cal B})$.  This property follows trivially from the definition of $\Sc_{\cal A}(\rho_{\cal A})$ and unitarity.  The analogous result $\chi_{\cal A} = \chi_{\cal B}$ also follows trivially from the definition of $\chi$.

\item \label{genprop:Max}
{\bf Coarse graining can only increase the von Neumann entropy:}  By virtue of the maximization condition in our definition of $\Sc_{\cal A}$
\begin{align}
\Sc_{\cal A}(\rho_{\cal A}) \ge S_{\cal A}(\rho_{\cal A}).
\end{align}
This property echoes the result of~\cite{Hubeny:2012wa,Wall:2012uf} that $\chi_{\cal A} \ge S_{\cal A}$.

\item \label{genprop:doubleCG}
{\bf The coarse-grained entropy is the entropy of the coarse-grained state:}  Given some state $\rho_{\cal A}$, if $\tau_{\cal A}$ is any state which satisfies the constraints~\eqref{eq:CGconstraint} (i.e. $\tau_{\cal A} \in T_{\cal A}(\rho_{\cal A})$) and $\sigma_{\cal A}$ is the coarse graining of $\rho_{\cal A}$ then
\begin{align}
  \Sc_{\cal A}(\rho_{\cal A}) = \Sc_{\cal A}(\tau_{\cal A}) =  \Sc_{\cal A}(\sigma_{\cal A}) = S_{\cal A}(\sigma_{\cal A}).
\end{align}

\end{Alist}

From these simple facts we learn two things.  First, if a coarse-grained entropy $\Sc$ is dual to $\chi$ then it must have the property that for any classical state $\rho_{\cal A}$
\begin{align} \label{eq:Areaconstraint}
\chi_{\cal A}(\rho_{\cal A}) = \chi_{\cal A}(\tau_{\cal A}),
\end{align}
where $\tau_{\cal A}$ is any other classical state in $T_{\cal A}(\rho_{\cal A})$.  We call any coarse graining which satisfies~\eqref{eq:Areaconstraint} a `$\chi$-preserving coarse graining.'    Second, if $\Sc$ is a $\chi$-preserving coarse-graining and $\rho_{\cal A}$ is a classical state for which the coarse-grained state $\sigma_{\cal A}$ is also classical then
\begin{align} \label{eq:Scinequality}
\Sc_{\cal A}(\rho_{\cal A}) \le \chi_{\cal A}(\rho_{\cal A}).
\end{align}

The conjunction of these results gives an even more useful result.  Let $\bar\Sc$ and $\tilde \Sc$ be two $\chi$-preserving coarse grainings and let $\bar\Sc \prec \tilde\Sc$.  Now let $\tilde R$ be the set of classical states which are mapped to classical coarse-grained states under the coarse graining $\tilde \Sc$.  We say that $\tilde\Sc$ is a `classical coarse graining' on $\tilde R$ and it follows that for any $\rho_{\cal A} \in \tilde R$
\begin{align} \label{eq:strongnessCondition}
\bar\Sc_{\cal A}(\rho_{\cal A}) \le \tilde\Sc_{\cal A}(\rho_{\cal A}) \le \chi_{\cal A}(\rho_{\cal A}).
\end{align}
This implies that $\bar\Sc$ cannot be dual to $\chi$ unless $\bar\Sc(\rho_{\cal A}) = \tilde\Sc(\rho_{\cal A})$ for all $\rho_{\cal A} \in \tilde R$.  In other words, if $\tilde\Sc$ is dual to $\chi$ it must be (at order $N^2$) as strong as possible over the states $\tilde R$.  This would imply that, up to equivalence, $\tilde \Sc$ would have to be the unique maximally-strong coarse graining over $\tilde R$, among those which are $\chi$-preserving and classical.

The restriction that $\tilde\Sc$ be as strong as possible only over the states $\tilde R$ is a little unwieldy since the definition of $\tilde R$ depends on $\tilde\Sc$.  So, it is natural to ask if the restriction to $\tilde R$ can simply be dropped, meaning that we would look for the strongest possible $\chi$-preserving coarse graining.  The answer is no, as we show in Appendix~\ref{app:chiPreserving}.  Given the importance of this restriction, it is interesting to consider $\chi$-preserving coarse grainings which map \emph{all} classical states to classical coarse-grained states.  (An example of such a coarse graining is the fine grained entropy $S$ which preserves the entire state.)  These completely classical coarse-grained entropies are particularly convenient to work with because in principle all of their properties can be derived by studying boundary value problems in classical general relativity.  While it is still logically consistent that $\chi$ is dual to a non-classical coarse graining, our intuition is that $\chi$ is dual to the strongest $\chi$-preserving coarse-grained entropy which always maps classical states to classical coarse-grained states.

In section~\ref{sec:OPentropy} we will define the one-point entropy $\ScOP$ and argue that it is the strongest, classical $\chi$-preserving coarse graining, at least in a particular perturbative context.

\section{The one-point entropy} \label{sec:OPentropy}

In this section we define a particular coarse-grained entropy which we call the `one-point entropy' $\ScOP$, and present evidence that it is dual to $\chi$ for theories without boundary sources (see appendix~\ref{app:BoundarySources}).  We will then compare the one-point entropy to other coarse-grained entropies, and indicate some potential future tests of our conjecture.

\subsection{Definition of the one-point entropy} \label{sec:OPdef}

The constraints $\{{\cal O}_m \}$ of $\ScOP_{\cal A}$ are the one-point functions of all gauge-invariant, local CFT operators supported on $D[{\cal A}]$.

Since we will only be testing our conjecture $\ScOP = \chi$ in the classical correspondence limit, many of the one-point CFT operators in $\{{\cal O}_m \}$ do not play much of a role.  This includes:
\begin{itemize}
\item Fermionic operators, because fermions anticommute and therefore it is difficult to make sense of them in the classical limit;
\item Multi-trace operators, because the asymptotic boundary values of the classical fields can be determined from the single-trace operators alone;
\item Operators whose dimension is parametrically large in N, because these correspond to very massive objects in the bulk, which are not contained in the classical supergravity field theory limit.
\end{itemize}
It is not clear to us whether operators like these should be included or excluded.  Possibly it makes no difference at order $N^2$, in which case either choice would lead to equivalent coarse grainings.\footnote{But one would have to make a definite choice if one tried to extend the conjecture to the semiclassical regime, as discussed in section~\ref{sec:Alt}.}  For the sake of definiteness, we define $\ScOP$ to include constraints from \emph{all} one-point functions.  However, the reader should bear in mind the other possibilities.

The AdS/CFT dictionary states that the single-trace one-point functions are given by
\begin{align} \label{eq:OnePointDef}
\left< {\cal O}_m (x) \right> = \frac{s}{\sqrt{-g}} \frac{\delta S_\textrm{ren} }{\delta \tilde\varphi(x)},
\end{align}
where $g$ is the determinant of the boundary metric $g_{\mu\nu}$, $\tilde\varphi$ is an appropriately conformally rescaled bulk field, $s$ is a conventional constant, and $S_\textrm{ren}$ is the renormalized action which includes the boundary counterterms required by the prescription of~\cite{Henningson:1998gx,Balasubramanian:1999re} (see~\cite{Skenderis:2002wp} for a review).  For example, the one-point functions of the stress tensor are given by
\begin{align}\label{eq:OnePointT}
\left< T^{\mu\nu}(x) \right> = \frac{2}{\sqrt{-g}} \frac{\delta S_\textrm{ren} }{\delta g_{\mu\nu}(x)},
\end{align}
with similar relations holding for all of the other bulk fields.  These relations allow us to express the constraints as a set of conditions on the asymptotic behavior of the bulk fields in $\bdiamond_{\cal A}$.

\subsection{Properties of the one-point entropy }

We now list some properties of the one-point entropy $\ScOP$ (beyond those in section~\ref{sec:genProp} which apply to all coarse grainings) that make it a promising candidate for the dual of $\chi$.

\begin{Blist}

\item \label{prop:factor}
{\bf The one-point entropy is additive for spacelike separated regions:}  Consider two spacelike separated boundary regions $\cal A$ and $\cal B$ for which $D[{\cal A}] \cap D[{\cal B}] = \emptyset$.  (Note that because these domains are closed, $D[{\cal A}]$ and $D[{\cal B}]$ cannot even touch at their boundaries.)  Consider the state $\rho_{\cal A} \otimes \rho_{\cal B}$.  This state is not in general the same state as $\rho_{{\cal A} \cup {\cal B}}$, because the correlations between A and B have been removed.  However, since the constraints~\eqref{eq:CGconstraint} only involve local operators, correlations between the two regions will not contribute to any of the expectation values of local operators, so the constraints factorize.  Thus, $\sigma_{{\cal A} \cup {\cal B}} = \sigma_{\cal A}\otimes \sigma_{\cal B}$ and we obtain
\begin{align}
\ScOP_{{\cal A} \cup {\cal B}} = \ScOP_{\cal A} + \ScOP_{\cal B}.
\end{align}

Now by boundary causality on the CFT, we know that there are no timelike or null causal curve connecting $D[{\cal A}]$ and $D[{\cal B}]$ in the bulk.  Hence the bulk causal wedges do not ``interact" and the causal holographic information obeys
\begin{align}
\chi_{{\cal A} \cup {\cal B}} = \chi_{\cal A} + \chi_{\cal B}.
\end{align}
A similar observation for a related proposal was previously made in \cite{Freivogel:2013zta} (see section~\ref{sec:Alt} for further discussion).

This is a special property of the one-point entropy.  A coarse-graining $\Sc^{(n)}$ which included the effects of higher n-point functions would not in general be additive, since it would be sensitive to correlations between two nearby regions ${\cal A}$ and ${\cal B}$.

\item \label{prop:pure}
{\bf The one-point entropy of a pure state does not always vanish:} Consider a thermal state $\rho_\textrm{thermal}$ with finite temperature $\beta > 0$.  A pure state $\left| \psi \right>$ for which
\begin{align}
\left< {\cal O}_m \right>_{\left| \psi \right> \left< \psi \right|} =\left< {\cal O}_m \right>_{\rho_\textrm{thermal}} ,
\end{align}
will have the property that for any Cauchy surface $\cal C$ we have $\ScOP_{\cal C}(\left| \psi \right> \left< \psi \right|) > 0$.  Note that we must use the limiting procedure described in section~\ref{sec:CorrPrinc} to exclude Poincar\'e recurrences or other large quantum fluctuations from our analysis.

\begin{figure}
\includegraphics[width=0.2 \textwidth]{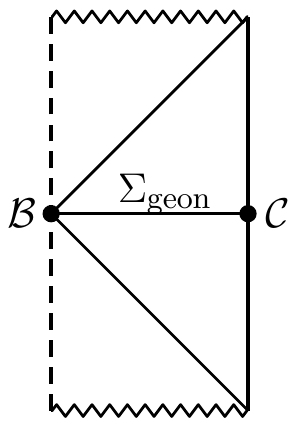} 
\caption{A causal diagram of the geon spacetime described in the text.  $\Sigma_\textrm{geon}$ is a bulk Cauchy surface, $\cal C$ is a boundary Cauchy surface and $\cal B$ is the bifurcation surface of the geon.}
\label{fig:geon}
\end{figure}

An interesting example of such states are topological geons~\cite{Sorkin:1986}.  The simplest geon solution is constructed by cutting off a $t=0$ slice of AdS-Schwarzschild at the bifurcation surface $\cal B$ and then identifying antipodal points on $\cal B$ to heal the geometry.  Call the resulting surface $\Sigma_\textrm{geon}$.  The maximal evolution of $\Sigma_\textrm{geon}$ is a spacetime that has AdS-Schwarzschild as its universal covering space (see Fig.~\ref{fig:geon}).  In $D=4$ spacetime dimensions this geometry is called a $\mathbb{RP}_3$ geon because its spatial slices have topology $\mathbb{RP}_3 - \{O\}$ where $O$ corresponds to spatial infinity (see e.g.~\cite{Friedman:1993ty}).

Now we will show that the CFT state $\rho_\textrm{geon}$ associated with this geometry is a pure state by calculating $S_{\cal C}(\rho_\textrm{geon})$, where $\cal C$ is a Cauchy surface of geon boundary.  The HRT proposal tells us that we must find the minimum-area extremal surface ${\mathfrak E}_{\cal C}$ that is homologous to $\cal C$.  As with AdS-Schwarzschild there are two candidate extremal surfaces: the empty set (with zero area) and the bifurcation surface (with finite area).  In AdS-Schwarzschild only the bifurcation surface is homologous to $\cal C$; therefore $S_{\cal C}(\rho_\textrm{thermal}) = S_{BH}$ (where $\rho_\textrm{thermal}$ is the dual CFT state and $S_{BH}$ is the Bekenstein-Hawking entropy).  But in the geon spacetime, the empty set is also homologous to $\cal C$; therefore $S_{\cal C}(\rho_\textrm{geon}) = 0$ (see also~\cite{Louko:1998hc}).

Next we calculate $\ScOP_{\cal C}(\rho_\textrm{geon})$.  By construction the geon spacetime is isometric to AdS-Schwarzschild in the exterior of the horizon.  It then follows trivially from the AdS/CFT dictionary~\eqref{eq:OnePointT} that the one-point functions of $\rho_\textrm{geon}$ and $\rho_\textrm{thermal}$ are equal.  Therefore, by~\ref{genprop:doubleCG} we have
\begin{align}
\ScOP_{\cal C}(\rho_\textrm{geon}) = S_{\cal C}(\rho_\textrm{thermal}) = S_{BH}.
\end{align}

Now on the bulk side, when we calculate $\chi_{\cal C}(\rho_\textrm{geon}) $ using the limiting procedure of~\eqref{eq:chiLimit} we also obtain $\chi_{\cal C} = S_{BH} = \ScOP_{\cal C}(\rho_\textrm{geon})$.  Again this follows trivially from the fact that the geon spacetime is isometric to AdS-Schwarzschild in the exterior of the horizon.\footnote{Note that had we not used~\eqref{eq:chiLimit} we would have incorrectly obtained $S_{BH}/2$ since the antipodal identification of the bifurcation surface effectively halves its area.  This quotient does not change the area of any other surface of the horizon, so the limit in~\eqref{eq:chiLimit} does not know about this discontinuity in the area.}  It is intriguing that this calculation relies crucially on the fact that $S$ depends on the global topology of the spacetime but $\chi$ does not.

The state $\rho_\textrm{geon}$ also provides an important counterexample useful for excluding coarse grainings weaker than $\ScOP$ (see section~\ref{sec:comparison} below).  We will now show that the states $\rho_\textrm{geon}$ and $\rho_\textrm{thermal}$ have different two-point functions.  Therefore a coarse graining $\Sc^{(2)}$ which constraints all one- and two-point function would have $\Sc^{(2)}(\rho_\textrm{geon}) < S_{BH}$ by~\eqref{eq:compent}.

Consider two points $x,y$ on the boundary of the geon spacetime.  In the free field limit, the two-point function is due to Witten diagrams which begin at $x$ and end at $y$ in position space.  Now because the geon is a quotient of AdS-Schwarzschild, it includes not only the Witten diagrams of AdS-Schwarzschild, but also noncontractable Witten diagrams which wrap around the nontrivial topology and make an additional contribution to the two-point function.  Therefore the two point functions of $\rho_\textrm{geon}$ and $\rho_\textrm{thermal}$ are not equal.\footnote{See~\cite{Smith:2013zqa} for explicit calculations showing that physical detectors placed outside of the horizon register the difference between the states $\rho_\textrm{geon}$ and $\rho_\textrm{thermal}$.}

\item \label{prop:complement}
{\bf For pure states, the one-point entropy of a region is generally not equal to the one-point region of the complementary region:}  This property follows immediately from~\ref{prop:pure} since for any Cauchy surface $\cal C$, $\ScOP_{{\cal C}^C} = 0$ but it was just shown that for some pure states $\ScOP_{{\cal C}} > 0$.  More generally if we take an arbitrary region $\cal A$ and act with an arbitrary unitary operator supported only in ${\cal A}^C$ we do not change $\ScOP_{{\cal A}}$, but will generally change $\ScOP_{{\cal A}^C}$ because the one-point functions are not invariant under unitary transformations.

Similarly, it was shown in~\cite{Hubeny:2012wa} (by applying the Gao-Wald focusing theorem~\cite{Gao:2000ga}) that generally $\chi_{\cal A} \ne \chi_{{\cal A}^C}$ for arbitrary regions $\cal A$.

\item \label{prop:ModHamiltonian}
{\bf The one-point entropy reduces to the fine-grained entropy for states which are thermal with respect to geometric flows:} This fact is of particular interest because Hubeny and Rangamani conjectured that $\chi_{\cal A} = S_{\cal A}$ if $\rho_{\cal A}$ is thermal~\cite{Hubeny:2012wa}.  By~\ref{lemma}, our proposal reproduces this result whenever the modular Hamiltonian (as defined in~\ref{lemma}) of $\rho$ is a linear combination of local operators.\footnote{See section~\ref{sec:comparison} for comparison with the results of~\cite{Freivogel:2013zta}. }  This happens to be true for all known cases in which $\chi_{\cal A} = S_{\cal A}$.  The known cases are
\begin{itemize}
\item
Spherical regions $\cal A$ in the vacuum state $\rho_\textrm{vacuum}$ of a CFT.  In this case the modular Hamiltonian of $\rho_{\cal A}$ is a diffeomorphism generator, and therefore a linear function of $T_{\mu \nu}$~\cite{Casini:2011kv}.

\item
Spherical regions $\cal A$ of the rotating BTZ geometry.  A change of coordinates maps the BTZ wedge $\bdiamond_{\cal A}$ onto a wedge to the AdS geometry and the previous argument applies.

\item
Certain eternal black holes (including charged and dilatonic black holes) are also dual to thermal states of the entire CFT.  The modular Hamiltonian is simply a linear combination of global charges of the spacetime and therefore $\ScOP_{\cal C} = S_{\cal C}  = S_{BH} = \chi_{\cal C}$, where $\cal C$ is a Cauchy surface.  (This shows that we need our coarse graining to constrain, not just the one-point function of the boundary stress-energy tensor $T_{\mu\nu}$, but also the CFT operators which are dual to the bulk dilaton and gauge fields.)

\end{itemize}

\item \label{prop:ThermalBound}
{\bf The one-point entropy is bounded by a thermal entropy:}  For any region $\cal A$
\begin{align} \label{eq:thermalMax}
\ScOP_{\cal A}(\rho_{\cal A}) \le S_{\cal A}(\rho_\textrm{thermal}) ,
\end{align}
where 
\begin{align}
\rho_\textrm{thermal} = Z^{-1} \exp(-\beta_\rho H).
\end{align}
In the previous expression $H \in \{ {\cal O}_m \}$ and $\beta_\rho$ is a constant chosen so that $\left<H\right>_{\rho_{\cal A}} = \left<H\right>_{\rho_\textrm{thermal}}$.

To see this note that $\rho_\textrm{thermal}$ maximizes the entropy subject to what amounts to a subset of the constraints~\eqref{eq:CGconstraint}, and imposing additional constraints cannot raise the entropy.  Furthermore, by~\ref{lemma} $ \ScOP_{\cal A}(\rho_\textrm{thermal}) = S_{\cal A}(\rho_\textrm{thermal})$ so we obtain~\eqref{eq:thermalMax}.

Now in the case of a Cauchy surface of an eternal black hole spacetime which is dual to a thermal state, the modular Hamiltonian $H$ is a linear combination of energy, angular momentum, and other global charges.  In this case,~\ref{prop:ModHamiltonian} implies that \eqref{eq:thermalMax} is saturated, so our proposal requires that black holes which are dual to thermal states always maximize their area subject to the constraint of fixed energy and other global charges.

\item \label{prop:acausalPert}
{\bf The one-point entropy is invariant under alterations to the dual spacetime outside the causal wedge:}  Consider some boundary region $\cal A$ with a classical reduced density matrix $\rho_{\cal A}$ dual to a bulk causal wedge $\bdiamond_{\cal A}$.  Now consider an alteration of the bulk spacetime which leaves the casual wedge of $\cal A$ unchanged, but which is not necessarily small anywhere else.  Such an alteration will produce a new reduced density matrix $\tau_{\cal A}$, which is in general \emph{not} equal to $\rho_{\cal A}$.  To see this, note that for generic spacetimes the extremal surface ${\mathfrak E}_{\cal A}$ lies outside of $\bdiamond_{\cal A}$~\cite{Hubeny:2012wa,Wall:2012uf}.  Therefore it is possible for a modification of the spacetime outside of $\bdiamond_{\cal A}$ to change the fine grained entropy, so that $S_{\cal A}(\tau_{\cal A}) \ne S_{\cal A}(\rho_{\cal A})$.  Now it follows immediately from the AdS/CFT dictionary~\eqref{eq:OnePointDef} and the locality of the bulk theory that any such perturbation will not change the one-point functions in $D[{\cal A}]$.  Therefore $\tau_{\cal A} \in T_{\cal A}(\rho_{\cal A})$, so $\ScOP_{\cal A}(\tau_{\cal A}) = \ScOP_{\cal A}(\rho_{\cal A})$.

By construction we have not modified the causal wedge $\bdiamond_{\cal A}$ so it immediately follows that $\chi_{\cal A}(\tau_{\cal A}) = \chi_{\cal A}(\rho_{\cal A})$.

\item \label{prop:BulkReco}
{\bf The one-point entropy is $\chi$-preserving in perturbation theory:}  Whereas~\ref{prop:acausalPert} showed that perturbations which do not alter $\bdiamond_{\cal A}$ (and therefore $\chi_{\cal A}$) preserve the one point functions, here we show a limited converse: that small perturbations which do not alter the one-point functions preserve $\bdiamond_{\cal A}$ and therefore $\chi_{\cal A}$.  

The problem of reconstructing the bulk given boundary data in asymptotically AdS spacetimes has been extensively studied~\cite{Balasubramanian:1998sn,Balasubramanian:1998de,Banks:1998dd,Bena:1999jv,Hamilton:2005ju,Hamilton:2006az,Kabat:2011rz,Heemskerk:2012mn}.  In the linearized bulk theory the boundary data in $\cal A$ is sufficient to reconstruct the fields in $\bdiamond_{\cal A}$; this construction can also be extended to the full nonlinear theory order-by-order in the interaction strength $\sqrt{G_N}$~\cite{Kabat:2011rz,Heemskerk:2012mn}.  In the correspondence limit, this boundary data reduces to one-point functions; therefore in the classical, perturbative regime, $\bdiamond_{\cal A}$ can be reconstructed from the one-point functions in $D[{\cal A}]$.

Now consider two states $\rho_{\cal A}$ and $\tau_{\cal A}$ which are perturbatively close to one another and have the same one-point functions.  Because they have the same one-point functions it follows immediately that $\ScOP_{\cal A}(\rho_{\cal A}) = \ScOP_{\cal A}(\tau_{\cal A})$.  Now in the bulk theory, the one-point functions completely determine the causal wedges associated with both states; therefore $\bdiamond_{\cal A}(\rho_{\cal A})=\bdiamond_{\cal A}(\tau_{\cal A})$ which implies $\chi_{\cal A}(\rho_{\cal A}) = \chi_{\cal A}(\tau_{\cal A})$.

\item \label{prop:collapse}
\begin{figure}
\includegraphics[width=0.2 \textwidth]{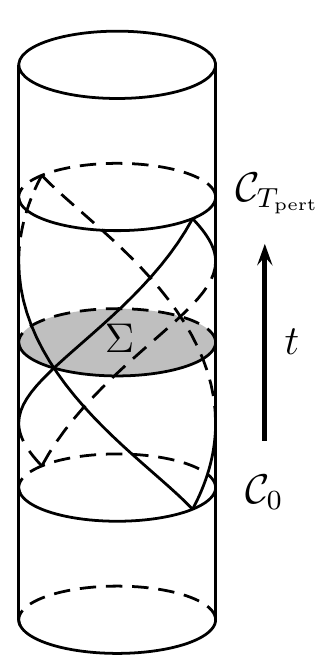} 
\caption{A sketch of the setup described in~\ref{prop:collapse}.  The spacetime is perturbatively close to vacuum AdS for a sufficiently long time $T_\textrm{pert}$ that a bulk Cauchy surface $\Sigma$ can be reconstructed from the boundary one-point functions.}
\label{fig:cauchy}
\end{figure}

{\bf The one-point entropy of a Cauchy surface vanishes for certain collapsed black holes:}  Consider a classical spacetime which is perturbatively close to vacuum AdS for a time $0 \le t \le T_\textrm{pert}$.  Let ${\cal C}_t$ be a family of boundary Cauchy surfaces and let $\cal M$ be the boundary region between ${\cal C}_0$ and ${\cal C}_{T_\textrm{pert}}$.  Let $T_\textrm{pert}$ be large enough that $J^+_\textrm{bulk}[{\cal M}] \cap J^-_\textrm{bulk}[{\cal M}]$ contains a bulk Cauchy surface $\Sigma$ (see Fig.~\ref{fig:cauchy}).  Let the set of all such states be called $R_{\chi=0}$.  The reconstruction results explained in~\ref{prop:BulkReco} imply that the classical Cauchy data on $\Sigma$ (and therefore the entire bulk spacetime) can be reconstructed from the boundary one-point functions in $\cal M$.\footnote{Note that by invoking~\ref{prop:BulkReco} we are implicitly assuming that the coarse grained state is perturbatively close to original state.  This seems plausible at least for \emph{some} class of small perturbations.}  Thus, the one-point entropy $\ScOP_{{\cal C}_t}(\rho_{{\cal C}_t})$ counts all states which correspond to this bulk geometry in the correspondence limit.  This quantity is precisely what is calculated by the Ryu-Takayanagi entropy $S_{{\cal C}_t}(\rho_{{\cal C}_t})$ so\footnote{Recall from section~\ref{sec:CorrPrinc} that we are only interested in the order $N^2$ pieces of $S$ and $\ScOP$.}
\begin{align} \label{eq:ScVac}
\ScOP_{{\cal C}_t}(\rho_{{\cal C}_t}) = S_{{\cal C}_t}(\rho_{{\cal C}_t}) = 0.
\end{align}
Now, by construction $\partial_+ \bdiamond_{{\cal C}_t}$ and $\partial_-\bdiamond_{{\cal C}_t}$ do not intersect.  This means that $\chi_{{\cal C}_t} = 0$, and so
\begin{align}
\ScOP_{{\cal C}_t}(\rho_{{\cal C}_t}) = \chi_{{\cal C}_t} = 0.
\end{align}
In~\cite{Bizon:2011gg,Jalmuzna:2011qw,Dias:2011ss} it is shown that AdS is perturbatively unstable to black hole collapse.  Thus almost all of the solutions we have considered will become black holes at late times.  The physical interpretation of $\chi_{{\cal C}_t} = 0$ for these states is that the one-point entropy is sensitive to the boundary data in the CFT, prior to the time that the state thermalizes.

\end{Blist}

\subsection{Comparison with other coarse-grained entropies} \label{sec:comparison}

We begin this section by showing that for the class of perturbative states $R_{\chi=0}$ considered in~\ref{prop:collapse}, $\ScOP$ is the strongest, classical $\chi$-preserving coarse grained entropy.  The key feature of the states $R_{\chi=0}$ are i) that there is a one-to-one map between boundary one-point functions and bulk causal wedges $\bdiamond_{{\cal C}_t}$ and ii) that $S_{{\cal C}_t} = 0 = \chi_{{\cal C}_t}$.

Since each classical state in $R_{\chi=0}$ is its own coarse graining, it follows that $\ScOP$ is $\chi$-preserving and classical over $R_{\chi=0}$.  Next, consider a stronger $\chi$-preserving coarse graining $\tilde\Sc \succ \ScOP$.  If $\tilde \Sc \not\equiv \ScOP$ then there must exist at least $O(N^2)$ classically distinguishable bulk wedges $\bdiamond^{(i)}$ 
that satisfy the constraints of $\tilde\Sc$ for some classical state $\rho_{{\cal C}_t}$.  All of these causal wedges have the same (vanishing) von Neumann entropy by the inequality $S_{{\cal C}_t} \le \chi_{{\cal C}_t} = 0$, therefore the coarse-grained state $\sigma_{{\cal C}_t}$ must be a mixture of the states dual to the $\bdiamond^{(i)}$. 
In other words, $\sigma_{{\cal C}_t}$ is not classical and so $\tilde\Sc$ is not classical over $R_{\chi=0}$.  Therefore there is no stronger, classical $\chi$-preserving coarse graining than $\ScOP$ over the states $R_{\chi=0}$.

Note that by~\ref{prop:acausalPert} and~\ref{prop:BulkReco}, $\ScOP$ is also $\chi$-preserving and classical in the perturbative regime for states with $\chi > 0$.  However, it is no longer trivial to show that any stronger $\chi$-preserving coarse graining is nonclassical.  Still, we conjecture that the obstacles to extending our argument are technical and that in fact $\ScOP$ is the strongest such coarse graining in this perturbative regime (in which we maximize entropy subject to the assumption that $\sigma$ is perturbatively close to $\rho$).

Throwing all caution to the winds, we conjecture that $\ScOP$ continues to be the strongest classical $\chi$-preserving coarse graining non-perturbatively.  
One can explore this question in classical general relativity, by asking if the bulk reconstruction results discussed in~\ref{prop:BulkReco} extend to the non-perturbative regime.  If not, it seems likely that the one-point functions do not fix $\chi$, in which case our conjecture $\ScOP = \chi$ can only work perturbatively.  In this case, it would be of interest to attempt to construct the strongest, classical $\chi$-preserving coarse graining explicitly (if it exists) and see if it is a candidate for the dual of $\chi$.

So, since we are not certain that $\ScOP$ is classical and $\chi$-preserving, it is worth considering if any weaker coarse graining might be viable.  One possibility is to consider a coarse-grained entropy $\Sc^{(2)} \prec \ScOP$ which constrains all one- and two-point functions.  However, we can show that $\Sc^{(2)}$ is inconsistent with the additivity property~\ref{prop:factor}.  Let $\cal A$ and $\cal B$ be two spherical regions on the vacuum AdS boundary, separated by a small spacelike gap.  For such regions the fine-grained entropy is subadditive: $S_{\cal A \cup B} \le S_{\cal A} + S_{\cal B}$.

By~\ref{prop:ModHamiltonian} we know that $\ScOP_{\cal A}(\rho_{\cal A}) = \Sc^{(2)}_{\cal A}(\rho_{\cal A}) = S_{\cal A}(\rho_{\cal A})$ and similarly for $\cal B$.  However, the two-point functions connecting regions $\cal A$ and $\cal B$ do not vanish, therefore $\sigma^{(1)}_{{\cal A}\cup{\cal B}} \not\in T^{(2)}_{{\cal A}\cup{\cal B}}(\rho_{{\cal A}\cup{\cal B}})$ (see~\ref{prop:factor}).  So, by~\eqref{eq:compent} we have
\begin{align}
\Sc^{(2)}_{{\cal A}\cup{\cal B}}(\rho_{{\cal A}\cup{\cal B}}) < \Sc^{(2)}_{\cal A}(\rho_{\cal A}) + \Sc^{(2)}_{\cal B}(\rho_{\cal B}).
\end{align}
Since the fine-grained entropy is subadditive at order $N^2$ we presume that $\Sc^{(2)}$ is as well.

One could try to evade this problem by strengthening $\Sc^{(2)}$.  Consider a coarse graining $\Sc^{(2\lozenge)}$ which constrains all one-point functions and those two-point functions for which both points are causally connected (c.f.~\cite{Balasubramanian:2013lsa}).  Now,  $\Sc^{(2\lozenge)}$ manifestly satisfies the additivity property~\ref{prop:factor}.  However, consider the states $\rho_\textrm{geon}$ and $\rho_\textrm{thermal}$ discussed in~\ref{prop:pure}.  These states have the same one-point functions but different two-point functions, therefore, $\rho_\textrm{thermal} \not\in T^{(2\lozenge)}(\rho_\textrm{geon})$.  It then follows from~\eqref{eq:compent} that for a Cauchy surface $\cal C$
\begin{align}
\Sc^{(2\lozenge)}_{\cal C}(\rho_\textrm{geon}) < \ScOP_{\cal C}(\rho_\textrm{geon}) = \chi_{\cal C}.
\end{align}
Assuming as above that this difference is of order $N^2$, this rules out 
$\Sc^{(2\lozenge)}$ and any weaker coarse graining as the dual of $\chi$.

Another conceivable weaker coarse graining might constrain all of the one-point functions and all Wilson loops.  However, Wilson loops are dual to extremal surfaces in the bulk geometry~\cite{Maldacena:1998im,*Rey:1998ik} and extremal surfaces can lie outside of $\bdiamond_{\cal A}$~\cite{Hubeny:2012ry}, in obvious tension with~\ref{prop:acausalPert}.\footnote{On the other hand, it has been argued~\cite{Louko:2000tp,Giddings:2001pt} that this duality is only valid in appropriately analytic spacetimes, and therefore it is not straightforward to draw inferences about causality.  So, this tension might have a resolution.}

It is also conceivable that some incomparable coarse graining $\hat\Sc \parallel \ScOP$ that combines partial data about the one-point functions and partial data about more complicated operators produce a candidate for the dual of $\chi$.  However, this type of construction seems likely to suffer from at least some of the shortcomings of both the stronger and weaker coarse grainings considered above.

Freivogel and Mosk have put forward a different kind of proposal for the dual of $\chi$~\cite{Freivogel:2013zta}.  Let $D[{\cal A}]$ be a simple causal diamond (i.e. it takes the form $J^{-}(p) \cap J^+(q)$ where $p$ and $q$ are points) on a conformally flat boundary metric.  The region $D[{\cal A}]$ thus has a time-translation conformal Killing vector $\xi$.  Now let $U= \exp(-i H t)$ be the unitary operator corresponding to the flow with respect to $\xi$.  The proposal of~\cite{Freivogel:2013zta} is that for such regions, $\chi_{\cal A} = \tilde S_{\cal A}( \rho_{\cal A})$, where
\begin{align} \label{eq:projection}
\tilde S_{\cal A}( \rho_{\cal A}) =  S_{\cal A}\left( \sum_i P_i \rho_{\cal A} P_i \right),
\end{align}
and the $P_i$ above are projection operators onto the eigenbasis of the operator $H$.  If $\rho_{\cal A}$ is a thermal state with modular Hamiltonian $H$ then $\tilde S_{\cal A}( \rho_{\cal A}) = S_{\cal A}(\rho_{\cal A})$, which reproduces the result~\ref{prop:ModHamiltonian} above.  Note that the projection $P_i \rho_{\cal A} P_i$ removes all off diagonal elements in the $H$ basis, which makes the resulting state time independent.  This corresponds to a coarse graining in which the constraints $\{{\cal O}_m\}$ consist of all functions of $H$.

The projection~\eqref{eq:projection} is equivalent to taking a time average of the state $\rho_{\cal A}$, which we call $\bar\rho_{\cal A}$.  Unfortunately, this implies that it is \emph{not} dual to $\chi$.  For consider an out of equilibrium state $\rho_{\cal A}$ which eventually (for very early and late modular times $t$) settles to an equilibrium state.  Let us suppose that in the bulk dual, this area of the future horizon at late times is equal to $A_\mathrm{final}$, as is the area of the past horizon at early times.  By the second law of horizons, $\chi(\rho_{\cal A}) < A_\mathrm{final}/4G_N $.  But inside of $\bdiamond_{\cal A}$, the time average of this bulk state is a stationary horizon with area $A_\mathrm{final}$.  Hence $\chi(\bar\rho_{\cal A}) = A_\mathrm{final}/4G_N$, so $\chi(\rho_{\cal A}) < \chi(\bar\rho_{\cal A})$ and the coarse graining $\tilde S$ is not $\chi$-preserving.\footnote{We owe this argument to Don Marolf.}

\subsection{Possible tests of \texorpdfstring{$\ScOP = \chi$}{TEXT}} \label{sec:tests}

While there is a great deal of data describing the behavior of $\chi$ in complex circumstances (see~\cite{Hubeny:2013hz,Hubeny:2013gba}), $\ScOP$ seems to be much less amenable to numerical calculation.   To test the conjecture, one may wish to look for aspects of $\ScOP$ (such as its divergence structure) which may be easy to calculate.  

An even better strategy for testing $\ScOP = \chi$ might be to identify circumstances in which our conjecture can be tested entirely within general relativity.  If two solutions exist with the same one-point functions and different values of $\chi$, then this would show that $\ScOP$ is not $\chi$-preserving and therefore not the dual of $\chi$.  Since the one-point functions correspond to the asymptotic values of classical fields, this leads to predictions about the allowed spacetimes on the bulk side.

Below we list a few special regimes in which it might be particularly easy to construct tests of our conjecture.

\begin{Clist}
\item
{\bf Spherical symmetry:} One strategy for finding solutions with the same one-point data is to exploit Birkhoff's theorem, which states that any spherically symmetric solution to general relativity with compactly supported matter will have one-point functions which are identical to AdS-Schwarzschild.

Now it is certainly possible to construct initial data that is spherically symmetric and has compactly supported matter.  However, evolving such initial data will generally lead to radiation which will propagate to the AdS boundary in finite time.    If this radiation can be suppressed in such a way that the presence of some matter alters $\chi_{\cal A}$ but no radiation reaches $D[{\cal A}]$, such a spacetime would be a counterexample to our conjecture that $\ScOP = \chi$.  There are several no-go theorems in general relativity that forbid ``horizonless solitons" (see e.g.~\cite{Gibbons:2013tqa} and references therein); however because the radiation only needs to be suppressed for a finite time these theorems are not sufficient by themselves to protect our conjecture.

In particular it would be interesting to attempt to construct such a solution using branes which have vanishing back reaction on the spacetime in the $N\rightarrow \infty$ limit.\footnote{Another intriguing possibility would be to study the Coulomb branch solutions considered in~\cite{Giddings:1999zu,Chepelev:1999zt}.}  Even though it is possible to construct spherically symmetric branes in AdS these branes are still localized on the compact dimensions and therefore may radiate via Kaluza-Klein modes.

\item
{\bf Null shock waves:}  Another approach to constructing counterexamples is to study null shock waves which pass through $\bdiamond_{\cal A}$ but which do not have an endpoint on $D[{\cal A}]$.  In~\cite{Bousso:2012mh,*Leichenauer:2013kaa} it is shown that the effect of such shock waves on the boundary one-point functions is heavily suppressed.  Thus it may be possible to bound the change in $\ScOP$ caused by these shock waves and compare it with the associated change in $\chi$.

\item \label{test:ScClass}
{\bf Generic coarse grained states:} Consider a generic boundary region $\cal A$ and associated with a bulk causal wedge $\bdiamond_{\cal A}$.  By~\ref{prop:acausalPert} arbitrary perturbations outside of $\bdiamond_{\cal A}$ will not affect $\ScOP_{\cal A}$ or $\chi_{\cal A}$ but they will generically change $S_{\cal A}$.  Now, by~\cite{Wall:2012uf} we must have $S_{\cal A} < \chi_{\cal A}$ for smooth generic spacetimes satisfying the null energy condition.  However, if $\chi_{\cal A} - S_{\cal A}$ can be made arbitrarily small then continuity would imply that if $\ScOP$ is classical, then it is dual to $\chi$.

Another approach would be to construct non-smooth spacetimes for which $S_{\cal A} = \chi_{\cal A}$ exactly.  Such spacetimes are reminiscent of the ``disentangled" Rindler wedges considered in~\cite{Czech:2012be}.  There it was shown that the Rindler horizons become singular when the entanglement between the two regions is no longer maximal.  These disentangled wedges could serve as a model for more general coarse grained states.

\item
{\bf Comparing divergences:} Freivogel and Mosk~\cite{Freivogel:2013zta} have calculated the logarithmically divergent piece of $\chi_{\cal A}$ for arbitrary regions $\cal A$ on a flat boundary in $D=4$ spacetime dimension.  They find that this logarithmic divergence is universal (i.e. independent of the state and the regulator) and that it \emph{cannot} be expressed as an integral of local geometric boundary quantities.  This means that unlike $S_{\cal A}$, the divergent terms in $\chi_{\cal A}$ are not dominated by vacuum correlations.  A greater understanding of coarse-grained states could allow comparison between the divergences of $\ScOP$ and those of $\chi$.  (Note that if $\sigma$ is a classical state, it must generically be nonsmooth at the causal surface, as shown in~\ref{test:ScClass}.  It is not surprising therefore that its divergences might differ from that of $\rho$.)

\item \label{test:reflect}

\begin{figure}
\includegraphics[width=0.25\textwidth]{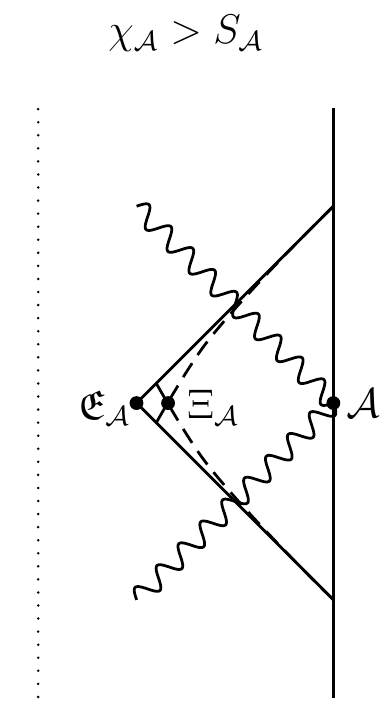} 
\caption{Matter reflecting off the AdS boundary.  The solid line to the right represents the AdS boundary and $\cal A$ is a spherical region (see~\ref{test:reflect}).}
\label{fig:reflect}
\end{figure}

{\bf Reflecting matter off the AdS boundary:}  Consider a spherical region $\cal A$ on the boundary of vacuum AdS.  The reduced density matrix associated with this region is the thermal state $\rho_{\cal A}$ (see~\ref{prop:ModHamiltonian}).  Now consider a state $\bar\rho_{\cal A} = e^{-i J} \rho_{\cal A} e^{i J}$ where $J$ is a source operator.  The spacetime associated with such a state will (for an appropriately chosen $J$) have a matter field bouncing off the AdS boundary (see Fig.~\ref{fig:reflect}).

Since the von Neumann entropy is preserved by unitary transformations and since $\rho_{\cal A}$ is thermal we know that $\ScOP(\bar\rho_{\cal A}) \ge \ScOP(\rho_{\cal A})$.  Furthermore $\tilde\rho_{\cal A}$ does not have the same one-point functions as $\rho_{\cal A}$ so it is unlikely that $\ScOP(\bar\rho_{\cal A}) = \ScOP(\rho_{\cal A})$ for general $U$.  Similarly, we know that $\chi_{\cal A}(\bar\rho_{\cal A}) > \ScOP(\rho_{\cal A})$.  It is conceivable that the state $\bar\rho_{\cal A}$ and its dual geometry could be constructed in sufficient detail to allow a precision test of $\ScOP = \chi$.

\item

{\bf Almost-complete Cauchy slices:}  Consider an eternal black hole in $D \ge 4$ spacetime dimensions and consider the quantity $\Delta S_{\cal A} = S_{\cal A}(\rho_{\cal A}) - S_{{\cal A}^C}(\rho_{{\cal A}^C})$.  It is well known that
\begin{align}
\lim_{{\cal A}^C \rightarrow \emptyset} \Delta S_{\cal A} = S_{BH},
\end{align}
and in fact $ \Delta S_{\cal A} = S_{BH}$ even when ${\cal A}^C$ is sufficiently small but finite.  In~\cite{Hubeny:2013gta} this leveling off of $\Delta S_{\cal A}$ is referred to as the entanglement plateaux.  

But for the causal surface, there is no plateaux.  If we now consider $\Delta \chi_{\cal A} = \chi_{\cal A} - \chi_{{\cal A}^C}$ we find that
\begin{align}
\lim_{{\cal A}^C \rightarrow \emptyset} \Delta \chi_{\cal A} > S_{BH},
\end{align}
even though \ref{prop:ModHamiltonian} says that $\chi_{\cal A} = S_{BH}$ when ${\cal A}^C = \emptyset$.  This means that  $\Delta \chi_{\cal A}$ jumps by a finite amount right when ${\cal A}$ becomes a complete Cauchy surface!  This effect is due to the red shift at the horizon, which prevents the causal surface from approaching arbitrarily close to the event horizon (Fig.~\ref{fig:BHandFunnel}(a)).  Can $\ScOP$ also jump in the same way (in the large N limit)?  If not, then our conjecture that $\ScOP_{\cal A}=\chi_{\cal A}$ would be falsified.  

\begin{figure}
\includegraphics[width=0.9 \textwidth]{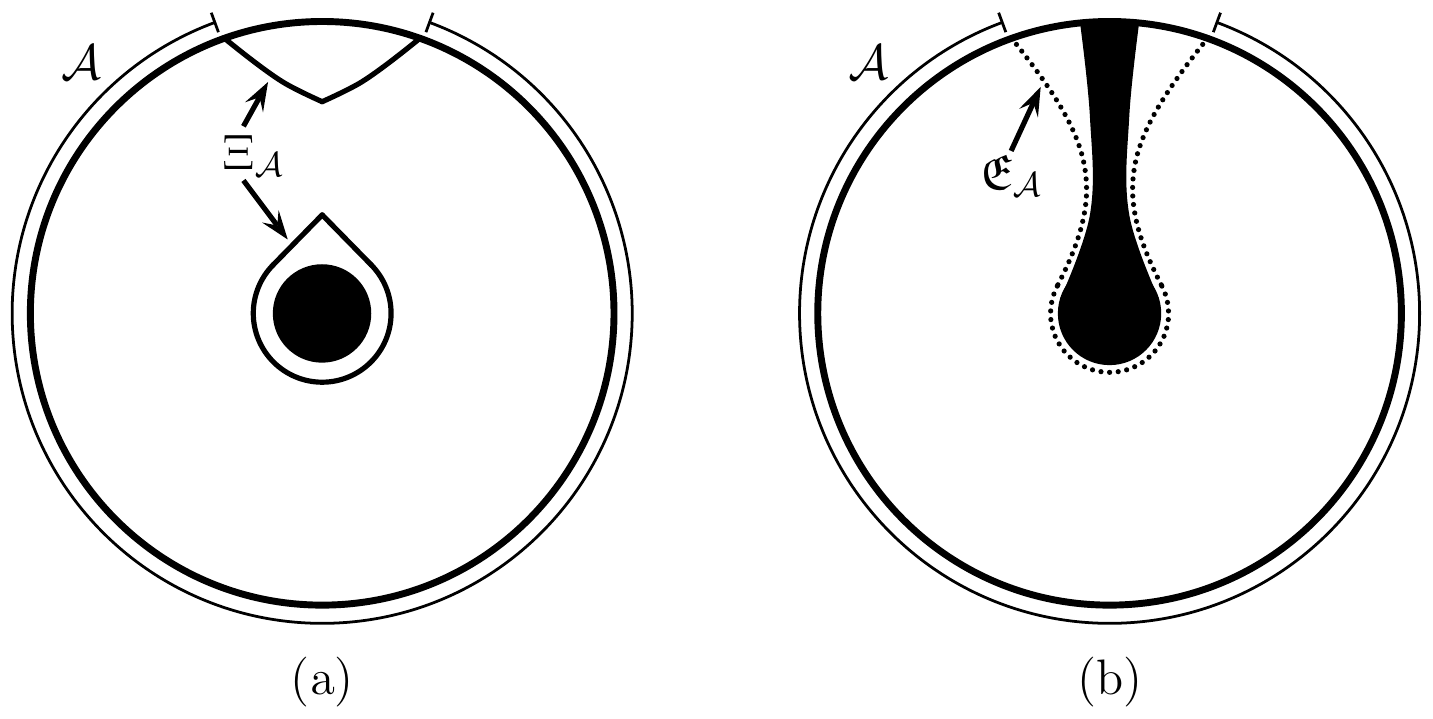} 
\caption{(a) A sketch of a $t=\textrm{constant}$ slice of the AdS-Schwarzschild solution.  Even for very small ${\cal A}^C$, $\chi_{\cal A}$ does not approach $S_{BH}$.  See~\cite{Hubeny:2013gba} for a precise diagram. (b) A bulk Cauchy surface of the (non-stationary) black funnel-like geometry discussed in the text.  The reduced density matrix on $\cal A$ is a candidate for a coarse graining of $\rho_{\cal A}$.}
\label{fig:BHandFunnel}
\end{figure}

Our conjecture requires that for arbitrarily small but finite ${\cal A}^C$, there must exist a state $\sigma_{\cal A}$ in $\cal A$ that has the same stress tensor $T_{\mu\nu}$ as the eternal black hole, and has entropy $S_{\cal A}(\sigma_{\cal A})= \ScOP_{\cal A}(\rho_{\cal A})$.  If we assume that $\ScOP$ is classical, then we can look for such states entirely within classical general relativity.  An interesting candidate state can be constructed by patching the region ${\cal A}$ to a Schwarzschild black hole.  Consider such a state with a time reflection symmetry on a Cauchy surface ${\cal C}$ which contains $\cal A$.  The horizon of this boundary black hole will extend into the bulk in a manner which might resemble a non-stationary black funnel-like spacetime sketched in Fig.~\ref{fig:BHandFunnel}(b) (see~\cite{Hubeny:2009ru,*Hubeny:2009kz,*Hubeny:2009rc,Santos:2012he}).\footnote{This solution can only exist if the one-point functions do not uniquely specify $\bdiamond_{\cal A}$ non-perturbatively.  Another interesting candidate for $\sigma_{\cal A}$ is the related black droplet solution.}  As noted in~\ref{test:ScClass}, $\sigma_{\cal A}$ cannot be smooth, however, it is possible that the required patching of the black hole disrupts the smoothness of the bulk geometry.  If it could be shown that such a solution exists and has $S_{\cal A}(\sigma_{\cal A}) =\Sc_{\cal A}(\sigma_{\cal A}) = \ScOP_{\cal A}(\rho_{\cal A})$ this would provide a nontrivial check on our proposal.

\end{Clist}

\section{The future one-point entropy} \label{sec:StrongCG}

\subsection{Motivation and definition}

Consider a pure state in AdS which, after some time, collapses to a black hole and rings down.  The HRT proposal assigns such a state zero entropy even at arbitrarily late times.  It is appropriate that a fine-grained notion of entropy should assign such a state zero entropy since the initial state is pure, and unitary evolution does not alter the entropy.  However, since this state is asymptotically stationary, at late times it is externally indistinguishable from an eternal black hole, which has a nonzero Bekenstein-Hawking entropy.  It is therefore tempting to apply the HRT proposal to the eternal black hole geometry, in order to calculate an approximate coarse-grained entropy.

Returning to the collapsing geometry, not only does the HRT entropy vanish for a Cauchy surface ${\cal C}$, but so do $\chi_{\cal C} $ and $\ScOP_{\cal C}$ (at least in the cases considered in~\ref{prop:collapse}).  We attribute this to the fact that the domain $D[{\cal C}]$ over which we coarse grain extends far into the past into the pre-thermalization region, when the geometry could easily be distinguished from a black hole.  While this is all perfectly consistent, it is not typically what is meant by a coarse-grained entropy, since it does not allow for thermalization.

Another feature that $\ScOP$ lacks that we might expect from a coarse-grained entropy is an interesting second law.  Technically $\ScOP_{\cal A}$ satisfies a second law (just like $S_{\cal A}$), however only in the trivial sense that
\begin{align}
\partial_t \left( \ScOP_{{\cal A}_t} \right)= 0
\end{align}
where ${\cal A}_t$ is a foliation of $D[{\cal A}]$ parameterized by $t$.

Motivated by the above concerns, we propose a new set of bulk and boundary quantities which we call the `future causal information' $\ph_{\cal A}$ and the `future one-point entropy' $\Ssc_{\cal A}(\rho_{\cal A})$.  We define 
\begin{align} \label{eq:SscDef}
\Ssc_{\cal A}(\rho_{\cal A}) = { \sup_{\tau_{\cal A} \in T^+_{\cal A} }} \left[ S_{\cal A}(\tau_{\cal A}) \right]
\end{align}
where $T^+_{\cal A}$ is the set of all density matrices which satisfy the constraints
\begin{align} \label{eq:CGconstraint2}
\Tr[{\cal O}_m \rho_{\cal A}] = \Tr[{\cal O}_m \tau_{\cal A}]
\end{align}
where now the $\{ {\cal O}_m \}$ in \eqref{eq:CGconstraint2} are the set of all one-point functions of the fields with support only on $D^+[{\cal A}]$.

\begin{figure}
\includegraphics[width=0.3\textwidth]{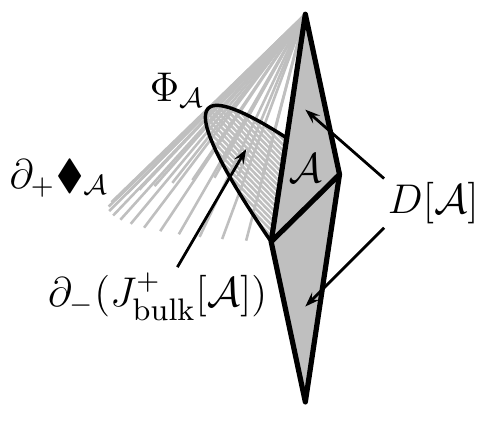} 
\caption{A sketch of the construction of $\Ph_{\cal A}$ described in the text.  $D[{\cal A}]$  is the boundary domain of dependence of $\cal A$ and $\Ph_{\cal A}$ extends into the bulk (see text).}
\label{fig:futureonepoint}
\end{figure}

We conjecture that in the absence of boundary sources, and in the correspondence limit of section~\ref{sec:CorrPrinc}, the bulk dual of $\Ssc_{\cal A}$ is given by 
\begin{align}
\Ssc_{\cal A} = \ph_{ A}  := \frac{\mathrm{Area}[\Ph_{\cal A}] }{4G},
\end{align}
where $\Ph_A$ is the codimension-two surface (see Fig.~\ref{fig:futureonepoint})
\begin{align} \label{eq:Phdef}
\Ph_{\cal A}  := \partial_+ \bdiamond_{\cal A} \cap \partial_- (J^+_\textrm{bulk}[{\cal A}]).
\end{align}

To summarize we have formed a new conjecture by modifying our old conjecture in two ways: the operators ${\cal O}_m$ are now supported on $D^+[{\cal A}]$ only as opposed to $D^+[{\cal A}] \cup D^-[{\cal A}]$, and the associated bulk surface is $\partial_+ \bdiamond_{\cal A} \cap \partial_- (J^+_\textrm{bulk}[{\cal A}])$ as opposed to $\partial_+ \bdiamond_{\cal A} \cap \partial_- \bdiamond_{\cal A}$.  We have again restricted our conjecture to theories without boundary sources for the reasons given in appendix~\ref{app:BoundarySources}.

\subsection{Properties of the future one-point entropy}

Note that lemma~\ref{lemma} and properties \ref{genprop:Max} and \ref{genprop:doubleCG} still apply to $\Ssc$.  However,~\ref{genprop:diamond} no longer applies, since $\Ssc$ now depends on the choice of $\cal A$, not just on $D[\cal A]$.  In addition $\Ssc$ has the following properties:

\begin{Dlist}

\item \label{propSsc:limits}
{\bf The future one-point entropy equals the one-point entropy if $\cal A$ is its own past:}  If ${\cal A} = D^-[{\cal A}]$ then $D^+[{\cal A}] = D[{\cal A}]$, and it follows that $\Ssc_{\cal A} = \ScOP_{\cal A}$.  In this case we also have $\ph_{\cal A} = \chi_{\cal A}$.  Thus if $\Ssc = \ph$ then it follows immediately that $\ScOP = \chi$. 

\item \label{propSsc:factor}
{\bf The future one-point entropy is additive for spacelike separated regions:}  Consider two spacelike separated regions $\cal A$ and $\cal B$ for which $D^+[{\cal A}] \cap D^+[{\cal B}] = \emptyset$.  Now if $D^+[{\cal A}] \cap D^+[{\cal B}] = \emptyset$ then it immediately follows that $D[{\cal A}] \cap D[{\cal B}] = \emptyset$.  Therefore, exactly as in~\ref{prop:factor}, we can consider the state $\rho_{\cal A}\otimes \rho_{\cal B}$ which differs from $\rho_{{\cal A} \cup {\cal B}}$ by correlations between $\cal A$ and $\cal B$.  Since the constraints are not sensitive to such correlations we obtain $\sigma_{{\cal A} \cup {\cal B}} = \sigma_{\cal A} \otimes \sigma_{\cal B}$ and
\begin{align}
\Ssc_{{\cal A} \cup {\cal B}} = \Ssc_{\cal A} + \Ssc_{\cal B}.
\end{align}

Since $D[{\cal A}] \cap D[{\cal B}] = \emptyset$, boundary causality requires that there are no bulk causal curves connecting $D^+[{\cal A}]$ and $D^+[{\cal B}]$; hence
\begin{align}
\ph_{{\cal A} \cup {\cal B}} = \ph_{\cal A} + \ph_{\cal B}.
\end{align}

\item \label{propSsc:2ndLaw}
{\bf The future one-point entropy obeys a non-trivial second law:}  Let $\cal A$ and $\cal B$ be two surfaces such that $D[{\cal A}] = D[{\cal B}]$ and let $\cal B$ lie nowhere to the past of $\cal A$.  Then 
\begin{align}
\Ssc_{\cal A} \le \Ssc_{\cal B}
\end{align}
due to the fact that the latter coarse graining has fewer constraints.

This matches the classical second law of causal horizons~\cite{Hawking:1971tu}, which says that for any causal horizon,
\begin{align}
\ph_{\cal A} \le \ph_{\cal B}.
\end{align}
In the case where $\cal C$ is a Cauchy surface, $\ph_{\cal C}$ corresponds to a slice of the global event horizon.  In the case where $D[{\cal A}]$ is a simple causal diamond, it corresponds to slices of an AdS-Rindler type causal horizon \cite{Jacobson:2003wv}.  In the most general case, it corresponds to the boundary of the past of some set of points $\cal Z$ on the AdS-boundary.  This is a slightly more general notion of causal horizon than that considered by \cite{Jacobson:2003wv} (which required the causal horizon to be the boundary of the past of a \emph{single} future-infinite worldline) but it still obeys a second law \cite{Wall:2010st}.

Note that although every choice of boundary slice $\cal B \in D[\cal A]$ maps to some slice $\phi_{\cal B}$ of the causal horizon, the map is neither one-to-one, nor onto.   If the null surface shot out from ${\cal B}$ develops caustics before intersecting the future horizon, then it is possible to modify parts of ${\cal B}$ without affecting $\phi_{\cal B}$.  Similarly, for any given slice $\phi$ there is no guarantee that there exists any dual choice of ${\cal B}$, since a null surface shot out from $\phi$ may also develop caustics.  Nevertheless it is remarkable that, if our conjecture is true, there exists an infinite-dimensional family of slices of the future horizon, whose (geometrical) bulk second law is dual to a (thermodynamic) boundary second law.

\item
{\bf The future one-point entropy is a stronger coarse graining than the one-point entropy:}  Since the maximization associated with $\Ssc_{\cal A}$ involves fewer constraints than that associated with $\ScOP_{\cal A}$, it follows that
\begin{align}
S \prec \ScOP \prec \Ssc,
\end{align}
where we have also used \ref{genprop:Max}.  Similarly from~\ref{propSsc:2ndLaw} we have
\begin{align}
S \le \chi \le \ph.
\end{align}

\item \label{propSsc:thermalization}
{\bf The future one-point entropy thermalizes:}  Let ${\cal C}_t$ be a foliation of Cauchy surfaces of a spacetime that starts as a small perturbation to AdS, but ultimately settles down to one or more black holes.  At early times, by~\ref{propSsc:limits}, we  recover
\begin{align}
\lim_{t\rightarrow -\infty} \Ssc_{{\cal C}_t}(\rho_{{\cal C}_t}) = \ScOP_{{\cal C}_t}(\rho_{{\cal C}_t}) = 0.
\end{align}
But at late times, the black holes ring down and the field theory state thermalizes.  In particular the one-point functions approach those of a thermal state, and we obtain
\begin{align}
\lim_{t \rightarrow \infty} \Ssc_{{\cal C}_t}(\rho_{{\cal A}_t}) = S_{{\cal C}_t}(\rho_\textrm{thermal}) = S_{BH}.
\end{align}
In the bulk geometry it follows from the causal structure of the spacetime that
\begin{align}
\lim_{t\rightarrow -\infty} \ph_{{\cal C}_t} = 0, \qquad \lim_{t\rightarrow \infty} \ph_{{\cal C}_t} = S_{BH}.
\end{align}
Again, we have used the limiting procedure of section~\ref{sec:CorrPrinc} to exclude Poincar\'e recurrences from our analysis.

There are also spacetimes which remain perturbatively close to AdS even at late times (see e.g.~\cite{Dias:2012tq}), for which $ \ph_{{\cal C}_t} =0$ for all $t$.  By the bulk reconstruction argument of~\ref{prop:collapse} these are precisely the state for which we would expect to have $ \Ssc_{{\cal C}_t} =0$ for all $t$ as well, since the entire bulk geometry can be reconstructed from one-point functions even at late times.

\item \label{propSsc:thermal}
{\bf The future one-point entropy reduces to the fine-grained entropy for states which are thermal with respect to geometric flows:}  By~\ref{prop:ModHamiltonian}, if $\cal A$ is a spherical region of the boundary of vacuum AdS, a BTZ black hole, or a Cauchy surface of an eternal black hole, then
\begin{align}
S_{\cal A}(\rho_{\cal A}) = \ScOP_{\cal A}(\rho_{\cal A}) = \Ssc_{\cal A}(\rho_{\cal A}).
\end{align}
This is also true for the associated bulk quantities even though $\Ph_{\cal A} \ne {\mathfrak E}_{\cal A} = \Xi_{\cal A}$.  This is because in each of these special cases, the future and past horizons of $D[{\cal A}]$ are stationary.  As a result, $\Ph_{\cal A}$ is connected to $\Xi_{\cal A}$ by a null congruence with zero expansion, so that $\chi_{\cal A} = \ph_{\cal A}$.

\item \label{propSsc:bound}
{\bf The future one-point entropy is bounded by a thermal entropy:}  Just as in~\ref{prop:ThermalBound}, for any region $\cal A$ if $\rho_\textrm{thermal}$ is a thermal state with modular Hamiltonian $H \in \{{\cal O}_m \}$ satisfying $\left<H\right>_{\rho_{\cal A}} = \left<H\right>_{\rho_\textrm{thermal}} $ then
\begin{align}
\Ssc_{\cal A}(\rho_{\cal A}) \le \Ssc_{\cal A}(\rho_\textrm{thermal}).
\end{align}
However, now we find that this bound is saturated not just by eternal black holes, but also by collapsed black holes in the limit that $\cal A$ sufficiently far to the future of the formation of the event horizon.

\end{Dlist}

It is worth emphasizing again that if our conjecture $\Ssc = \ph$ is correct, then the thermodynamic second law of $\Ssc$ of~\ref{propSsc:2ndLaw} is the bulk dual of the Hawking area increase theorem~\cite{Hawking:1971tu}, as applied to certain kinds of causal horizons~\cite{Jacobson:2003wv,Wall:2010st}.  In this way our proposal provides a quantum mechanical interpretation of the area law in terms of a thermodynamic second law in the boundary theory.

\subsection{Generalization to arbitrary boundary regions}

\begin{figure}
\includegraphics[width=0.3\textwidth]{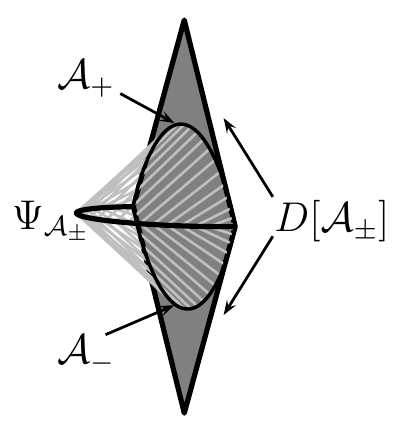} 
\caption{A sketch of the construction of $\Psi_{{\cal A}_-,{\cal A}_+}$ described in the text.  $D[{\cal A}_-] = D[{\cal A}_+]$  is the boundary domain of dependence of ${\cal A}_\pm$ and $\Psi_{{\cal A}_-,{\cal A}_+}$ extends into the bulk (see text).}
\label{fig:Psi}
\end{figure}

The generalization of $\chi$ to $\ph$ suggests a further generalization to more general bulk wedges.  Consider two regions ${\cal A}_-$ and ${\cal A}_+$ which have the same domain of dependence $D[{\cal A}_-] = D[{\cal A}_+]$ and for which ${\cal A}_+$  is everywhere to the future of ${\cal A}_-$, i.e. ${\cal A}_+ \in J^+[{\cal A}_-]$.  A natural generalization of~\eqref{eq:Phdef} is then to consider the surface (see Fig.~\ref{fig:Psi})
\begin{align}
\Psi_{{\cal A}_-,{\cal A}_+} = \partial_+ ( J^-_\textrm{bulk}[{\cal A}_+]) \cap \partial_- ( J^+_\textrm{bulk}[{\cal A}_-]).
\end{align}
Based on our previous experience it is tempting to conjecture that $\psi := \mathrm{Area}[\Psi]/4 G_N$ is dual to a coarse-grained entropy $\mathbb S^{(1)}$ whose constraints $\{{\cal O}_m\}$ are all one-point function supported in the region $J^+[{\cal A}_-] \cap J^-[{\cal A}_+]$.  However, this proposal meets with serious difficulties right away.

Let ${\cal C}_-$ and ${\cal C}_+$ be two Cauchy surfaces on the boundary of the AdS vacuum so that the region between ${\cal C}_-$ and ${\cal C}_+$ forms a strip.  The constraints associated with this strip include the total energy of the spacetime, which vanishes for vacuum AdS.  Since the AdS vacuum is the unique state in the theory with $E=0$, it follows that $\mathbb S^{(1)}_{{\cal C}_-,{\cal C}_+} = 0$ for any choice of ${\cal C}_-$ and ${\cal C}_+$.  Yet in the bulk, we have $\psi_{{\cal C}_-,{\cal C}_+} = 0$ only if ${\cal C}_-$ and ${\cal C}_+$ are separated by an AdS light crossing time or more.  Therefore, we find that $\psi_{{\cal C}_-,{\cal C}_+} > \mathbb S^{(1)}_{{\cal C}_-,{\cal C}_+}$ for certain choices of ${\cal C}_-,{\cal C}_+$.

It is hard to imagine how we might modify $\mathbb S^{(1)}$ in order to make a credible candidate for the dual of $\psi$.  One possibility is to introduce finite imprecision into the constraints, roughly as proposed in footnote~\ref{foot:approx}.  In particular we would need to the precision to depend on the width of the strip.  This is in some ways reminiscent of the Heisenberg uncertainty principle, which limits the precision with which the energy can be measured by coupling to a classical system for a finite time.  Bounds of this kind were found in the ``holographic thought experiments" of~\cite{Marolf:2008mg}.  However, it is unclear how to translate these ideas into a precise proposal for the dual of $\psi$.

A very different way of interpreting $\psi_{{\cal C}_-,{\cal C}_+}$ is put forward in~\cite{Balasubramanian:2013rqa,Balasubramanian:2013lsa}.  Balasubramanian et. al. propose that $\psi_{{\cal C}_-,{\cal C}_+}$ measures the entanglement between spatial regions separated by $\Psi_{{\cal C}_-,{\cal C}_+}$, which in the field theory roughly translates to entanglement between UV and IR degrees of freedom.  It would be very interesting to know if this entanglement entropy could be formulated as a  coarse-grained entropy which preserves the appropriate IR degrees of freedom.

\section{Discussion} \label{sec:Alt}

In summary, we have examined two coarse-grained entropies $\ScOP$ and $\Ssc$ in detail and found that they are plausibly dual to the causal holographic information $\chi$ and the future causal information $\ph$, respectively.  We have tested these conjectures by finding shared properties, and eliminating several classes of alternate proposals.

The evidence for our conjectures includes that i) both $\ScOP$ and $\Ssc$ are additive, as are their bulk duals (see~\ref{prop:factor},~\ref{propSsc:factor}), ii) $\ScOP = \chi$ and $\Ssc = \ph$ for thermal states and for the pure geon state (see~\ref{prop:pure},~\ref{prop:ModHamiltonian},~\ref{propSsc:thermal},~\ref{propSsc:bound}), and iii) in certain circumstances, the classical bulk spacetime can be reconstructed from the one-point functions (see~\ref{prop:BulkReco}), as discussed below.  Additionally, for the future one-point entropy, iv) $\Ssc$ obeys a second law (see~\ref{propSsc:2ndLaw}), and thermalizes in a way which correctly reproduces the early and late time entropy of a collapsing black hole (see~\ref{propSsc:thermalization}).

Assuming that the dual of $\chi$ is a member of a particularly nice class of coarse grainings, we can show that it must be the \emph{strongest} such coarse graining.  This class consists of those coarse-grainings which preserve $\chi$ and map classical states to classical states.  If the dual of $\chi$ belongs to this class, then (at least for these classical states) it must be the strongest possible such coarse graining, at order $N^2$.  In certain perturbative contexts, we have shown that $\ScOP$ does indeed belong to this class, and for the states $R_{\chi=0}$ considered in \ref{sec:comparison} we have also shown that it is the strongest.  Even for perturbations to geometries with $\chi > 0$, the bulk reconstruction theorems discussed in~\ref{prop:BulkReco} suggest that it is still the strongest.

Our conjecture is on more dubious ground non-perturbatively, but we have identified situations in which it can be tested using classical general relativity.  Several tests (some of which are non-perturbative) are listed in section~\ref{sec:tests}.  We believe that experts will be able to falsify or confirm our conjecture using existing analytic and numerical methods.

The most striking feature of $\Ssc$ is that it obeys a nontrivial second law (cf.~\ref{propSsc:2ndLaw}).  This allows us to describe the thermalization of CFT states, in a way which---if our conjecture is correct---is dual to the Hawking area theorem in the bulk.  However, the second law is a general feature of any coarse graining based on maximizing entropy subject to diminishing constraints.  So this property is not unique to the one-point constraints.  However the bulk reconstruction theorems tell us that the one-point entropy thermalizes in a way which is qualitatively similar to the collapse of a black hole as argued in~\ref{propSsc:thermalization}.

Finally we note that even though we have only analyzed the coarse-grained entropies $\ScOP$ and $\Ssc$ in the correspondence limit, these quantities are well defined at finite $N$, if one includes all local operators as prescribed in section~\ref{sec:OPdef}.  Are there still nice bulk duals for these quantities?

One can start by looking at the semiclassical regime.  In the boundary, this corresponds to taking the $N\to \infty$ limit, yet keeping terms subleading in $N$.  In this regime, the area of the HRT must be surface be corrected by adding a term which equal to the entanglement entropy across the surface \cite{Faulkner:2013ana}.  In other words, $S$ on the boundary is dual to the \emph{generalized entropy} of the HRT surface.

It is natural to suppose that $\chi$ and $\ph$ must be corrected in the same way.  Note that $\ph$ no longer obeys a second law because quantum matter fields can violate the null energy condition.  However, $\Ssc$ still obeys a second law, and so does the generalized entropy associated with $\ph$~\cite{Wall:2011hj}.  But unlike $\chi$ and $\ph$, the generalized entropy is not additive.  Perhaps this proposal can be saved by restricting to connected boundary regions, or by including higher-point functions at finite precision in $N$ (cf. footnote~\ref{foot:approx}).

\section*{Acknowledgements}

It is a pleasure to thank Bartek Czech, Sebastian Fischetti, Ben Freivogel, Gary Horowitz, Veronika Hubeny, Don Marolf, Mukund Rangamani, and Jorge Santos for useful discussions.  We also thank Veronika Hubeny, Don Marolf, and Mukund Rangamani for helpful comments on an early draft of this paper.  Additionally, WK thanks the organizers of the ``Gravity - New perspectives from strings and higher dimensions" workshop in Benasque, Spain for hospitality during the early stages of this project.  This work is supported in part by the National Science Foundation under Grant No PHY12-05500, and by funds from the University of California.  AW is also supported by the Simons Foundation.

\appendix

\section{\texorpdfstring{$\chi$}{TEXT}-preserving coarse grainings} \label{app:chiPreserving}

As mentioned in section~\ref{sec:genProp}, it is natural to ask if the restriction of \eqref{eq:strongnessCondition} to states with classical coarse grainings can be dropped.  In this appendix we show that the answer to this question is no.

Consider a coarse graining with the single constraint that $\left< \hat\chi \right>$ be held fixed, where $\left< \hat\chi \right>$ is some linear quantum expectation value which equals $\chi$ for classical states.  This coarse graining, which we call $\ScChi$, cannot be the dual of $\chi$.  Consider any Cauchy surface $\cal C$ and state $\rho_{\cal C}$ for which $\chi_{\cal C} = 0$.  The entropy $\ScChi_{\cal C}(\rho_{\cal C})$ counts all states for which $\chi_{\cal C} = 0$.  Because the volume of AdS is infinite, there are an infinite number of such states even at finite $N$.  Therefore $\ScChi_{\cal C}(\rho_{\cal C})$ diverges (beyond the usual $N^2$ divergence) in the correspondence limit.

$\Sc^{(\hat \chi)}$ is therefore pathological since it assigns infinite entropy to a pure state.  However, we can easily tame this divergence by adding a second constraint $\left< \int T_{tt} \right>$, which for a Cauchy surface ${\cal C}$ is simply the total energy $E$ of the spacetime.  Call this new coarse-grained entropy $\Sc^{(\hat \chi,E)}$.  Now the state counting for $\rho_{{\cal C}}$ includes all ways to collapse a black hole of a particular energy, including very slow collapses (e.g. the time reversal of Hawking evaporation for a sufficiently small black hole).  This quantity is finite but still of order $N^2$, which implies
\begin{align}
\Sc^{(\hat \chi,E)}_{{\cal C}}(\rho_{{\cal C}}) > \chi_{{\cal C}}.
\end{align}
We have not violated the inequality~\eqref{eq:strongnessCondition} because~\eqref{eq:strongnessCondition} only holds when the coarse-grained state $\sigma_{{\cal C}}$ is classical.  However, all of the classical states satisfying the constraints of $\Sc^{(\hat \chi,E)}$ have the same (vanishing) von Neumann entropy (since $S_{{\cal C}} \le \chi_{{\cal C}} = 0$ for all such classical geometries).  Hence the coarse graining $\sigma_{{\cal C}}$ is a mixture of an infinite number of classically distinguishable states, and therefore it is non-classical.

\section{Boundary sources} \label{app:BoundarySources}

As mentioned above, we only conjecture that $\chi$ is dual to a coarse-grained entropy for theories with time-independent Hamiltonians (i.e. in the absence of boundary sources).  We now explain the reason for this restriction.  

Let $\Sc$ be any coarse graining and let $\rho_{\cal A}$ be any state which satisfies the conditions of~\ref{lemma} so that $\Sc_{\cal A}(\rho_{\cal A}) = S_{\cal A}(\rho_{\cal A})$.  An important feature of~\ref{lemma} is that nothing is assumed about the time evolution of $\rho_{\cal A}$ within $D[{\cal A}]$, except that it is unitary.  It therefore applies even if we insert boundary sources, which can potentially increase $\chi_{\cal A}$.

\begin{figure}[h]
\includegraphics[width=0.5\textwidth]{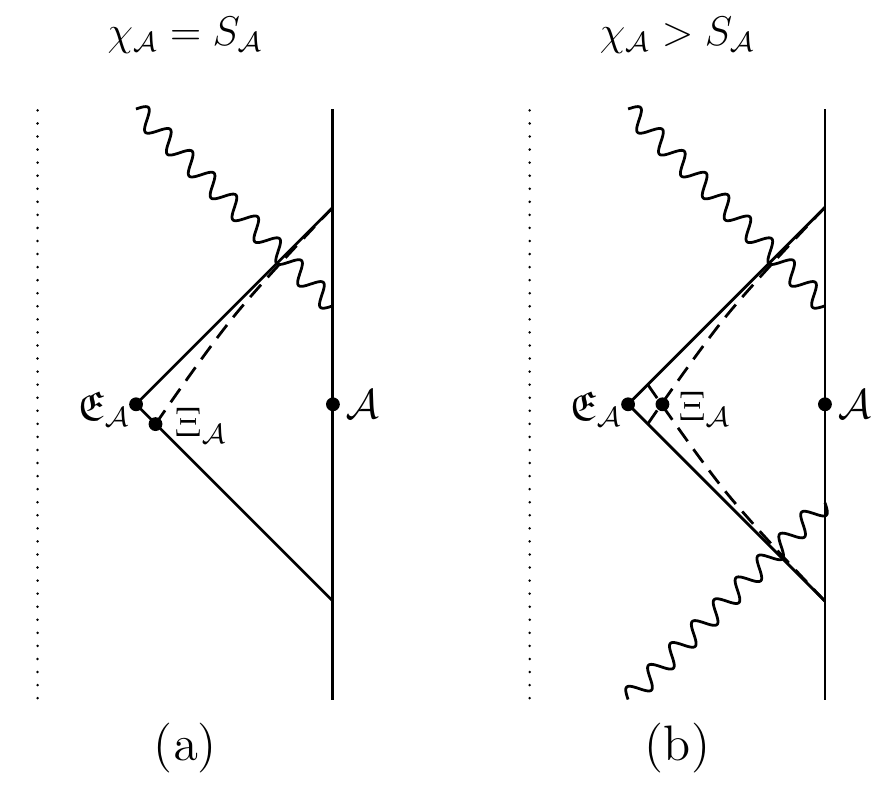} 
\caption{Various insertions of sources on the vacuum AdS boundary.  In each figure the solid line to the right represents the AdS boundary and $\cal A$ is a spherical region.  (a) By causality ${\mathfrak E}_{\cal A}$ is unperturbed by the sources however $\Xi_{\cal A}$ is moved due to focusing of light rays (shown schematically by the dashed lines).  However this focusing does not change $\chi$ since the past horizon has vanishing expansion.  (b) An ingoing and outgoing source which gives $\chi_{\cal A} > S_{\cal A} = \ScOP_{\cal A}$.}
\label{fig:source}
\end{figure}

This would lead to a contradiction in situations where $H \in \{ {\cal O}_m \}$, since we can always add or remove boundary sources to achieve $\Sc_{\cal A}(\rho_{\cal A}) < \chi_{\cal A}(\rho_{\cal A})$ (see Fig.~\ref{fig:source}).

This includes the case in which $\cal A$ is a Cauchy surface and the bulk geometry is a stationary black hole.  In this case the modular Hamiltonian is a linear combination of energy, angular momentum, gauge charges, etc.   It is hard to imagine a $\chi$-preserving coarse graining which does not constrain any of these quantities, and yet which does not suffer from the same problems as $\Sc^{(\hat \chi,E)}$ (see appendix~\ref{app:chiPreserving}).  For this reason we will restrict our attention to theories without any boundary sources turned on.

\bibliographystyle{kp}

\bibliography{ProposalJHEPv2.bbl}

\begingroup\raggedright\begin{thebibliography}{88}
\expandafter\ifx\csname natexlab\endcsname\relax\def\natexlab#1{#1}\fi

\bibitem[Maldacena(1998)]{Maldacena:1997re}
J.~M. Maldacena, ``{The Large N limit of superconformal field theories and
  supergravity}'', {\em Adv.Theor.Math.Phys.} {\bfseries 2} (1998) 231--252,
 \href{http://xxx.lanl.gov/abs/hep-th/9711200}{{\ttfamily
  arXiv:hep-th/9711200}}.

\bibitem[Gubser et~al.(1998)Gubser, Klebanov, and Polyakov]{Gubser:1998bc}
S.~Gubser, I.~R. Klebanov, and A.~M. Polyakov, ``{Gauge theory correlators from
  noncritical string theory}'', {\em Phys.Lett.} {\bfseries B428} (1998)
  105--114,
 \href{http://xxx.lanl.gov/abs/hep-th/9802109}{{\ttfamily
  arXiv:hep-th/9802109}}.

\bibitem[Witten(1998)]{Witten:1998qj}
E.~Witten, ``{Anti-de Sitter space and holography}'', {\em
  Adv.Theor.Math.Phys.} {\bfseries 2} (1998) 253--291,
 \href{http://xxx.lanl.gov/abs/hep-th/9802150}{{\ttfamily
  arXiv:hep-th/9802150}}.

\bibitem[Swingle(2012)]{Swingle:2009bg}
B.~Swingle, ``{Entanglement Renormalization and Holography}'', {\em Phys.Rev.}
  {\bfseries D86} (2012) 065007,
 \href{http://xxx.lanl.gov/abs/0905.1317}{{\ttfamily arXiv:0905.1317}}.

\bibitem[Heemskerk and Polchinski(2011)]{Heemskerk:2010hk}
I.~Heemskerk and J.~Polchinski, ``{Holographic and Wilsonian Renormalization
  Groups}'', {\em JHEP} {\bfseries 1106} (2011) 031,
 \href{http://xxx.lanl.gov/abs/1010.1264}{{\ttfamily arXiv:1010.1264}}.

\bibitem[Faulkner et~al.(2011)Faulkner, Liu, and Rangamani]{Faulkner:2010jy}
T.~Faulkner, H.~Liu, and M.~Rangamani, ``{Integrating out geometry: Holographic
  Wilsonian RG and the membrane paradigm}'', {\em JHEP} {\bfseries 1108} (2011)
  051,
 \href{http://xxx.lanl.gov/abs/1010.4036}{{\ttfamily arXiv:1010.4036}}.

\bibitem[Lee(2013)]{Lee:2013dln}
S.-S. Lee, ``{Quantum Renormalization Group and Holography}'',
 \href{http://xxx.lanl.gov/abs/1305.3908}{{\ttfamily arXiv:1305.3908}}.

\bibitem[Pauli and Villars(1949)]{Pauli:1949zm}
W.~Pauli and F.~Villars, ``{On the Invariant regularization in relativistic
  quantum theory}'', {\em Rev.Mod.Phys.} {\bfseries 21} (1949)
434--444.

\bibitem[Ryu and Takayanagi(2006{\natexlab{a}})]{Ryu:2006bv}
S.~Ryu and T.~Takayanagi, ``{Holographic derivation of entanglement entropy
  from AdS/CFT}'', {\em Phys.Rev.Lett.} {\bfseries 96} (2006){\natexlab{a}}
  181602,
 \href{http://xxx.lanl.gov/abs/hep-th/0603001}{{\ttfamily
  arXiv:hep-th/0603001}}.

\bibitem[Ryu and Takayanagi(2006{\natexlab{b}})]{Ryu:2006ef}
S.~Ryu and T.~Takayanagi, ``{Aspects of Holographic Entanglement Entropy}'',
  {\em JHEP} {\bfseries 0608} (2006){\natexlab{b}} 045,
 \href{http://xxx.lanl.gov/abs/hep-th/0605073}{{\ttfamily
  arXiv:hep-th/0605073}}.

\bibitem[Hubeny et~al.(2007)Hubeny, Rangamani, and Takayanagi]{Hubeny:2007xt}
V.~E. Hubeny, M.~Rangamani, and T.~Takayanagi, ``{A Covariant holographic
  entanglement entropy proposal}'', {\em JHEP} {\bfseries 0707} (2007) 062,
 \href{http://xxx.lanl.gov/abs/0705.0016}{{\ttfamily arXiv:0705.0016}}.

\bibitem[Wall(2012)]{Wall:2012uf}
A.~C. Wall, ``{Maximin Surfaces, and the Strong Subadditivity of the Covariant
  Holographic Entanglement Entropy}'',
 \href{http://xxx.lanl.gov/abs/1211.3494}{{\ttfamily arXiv:1211.3494}}.

\bibitem[Fursaev(2006)]{Fursaev:2006ih}
D.~V. Fursaev, ``{Proof of the holographic formula for entanglement entropy}'',
  {\em JHEP} {\bfseries 0609} (2006) 018,
 \href{http://xxx.lanl.gov/abs/hep-th/0606184}{{\ttfamily
  arXiv:hep-th/0606184}}.

\bibitem[Headrick(2010)]{Headrick:2010zt}
M.~Headrick, ``{Entanglement Renyi entropies in holographic theories}'', {\em
  Phys.Rev.} {\bfseries D82} (2010) 126010,
 \href{http://xxx.lanl.gov/abs/1006.0047}{{\ttfamily arXiv:1006.0047}}.

\bibitem[Casini et~al.(2011)Casini, Huerta, and Myers]{Casini:2011kv}
H.~Casini, M.~Huerta, and R.~C. Myers, ``{Towards a derivation of holographic
  entanglement entropy}'', {\em JHEP} {\bfseries 1105} (2011) 036,
 \href{http://xxx.lanl.gov/abs/1102.0440}{{\ttfamily arXiv:1102.0440}}.

\bibitem[Lewkowycz and Maldacena(2013)]{Lewkowycz:2013nqa}
A.~Lewkowycz and J.~Maldacena, ``{Generalized gravitational entropy}'',
 \href{http://xxx.lanl.gov/abs/1304.4926}{{\ttfamily arXiv:1304.4926}}.

\bibitem[Hartman(2013)]{Hartman:2013mia}
T.~Hartman, ``{Entanglement Entropy at Large Central Charge}'',
 \href{http://xxx.lanl.gov/abs/1303.6955}{{\ttfamily arXiv:1303.6955}}.

\bibitem[Faulkner(2013)]{Faulkner:2013yia}
T.~Faulkner, ``{The Entanglement Renyi Entropies of Disjoint Intervals in
  AdS/CFT}'',
 \href{http://xxx.lanl.gov/abs/1303.7221}{{\ttfamily arXiv:1303.7221}}.

\bibitem[Hammersley(2006)]{Hammersley:2006cp}
J.~Hammersley, ``{Extracting the bulk metric from boundary information in
  asymptotically AdS spacetimes}'', {\em JHEP} {\bfseries 0612} (2006) 047,
 \href{http://xxx.lanl.gov/abs/hep-th/0609202}{{\ttfamily
  arXiv:hep-th/0609202}}.

\bibitem[Hammersley(2008)]{Hammersley:2007ab}
J.~Hammersley, ``{Numerical metric extraction in AdS/CFT}'', {\em
  Gen.Rel.Grav.} {\bfseries 40} (2008) 1619--1652,
 \href{http://xxx.lanl.gov/abs/0705.0159}{{\ttfamily arXiv:0705.0159}}.

\bibitem[Bilson(2008)]{Bilson:2008ab}
S.~Bilson, ``{Extracting spacetimes using the AdS/CFT conjecture}'', {\em JHEP}
  {\bfseries 0808} (2008) 073,
 \href{http://xxx.lanl.gov/abs/0807.3695}{{\ttfamily arXiv:0807.3695}}.

\bibitem[Nozaki et~al.(2013)Nozaki, Numasawa, Prudenziati, and
  Takayanagi]{Nozaki:2013vta}
M.~Nozaki, T.~Numasawa, A.~Prudenziati, and T.~Takayanagi, ``{Dynamics of
  Entanglement Entropy from Einstein Equation}'', {\em Phys.Rev.} {\bfseries
  D88} (2013) 026012,
 \href{http://xxx.lanl.gov/abs/1304.7100}{{\ttfamily arXiv:1304.7100}}.

\bibitem[Lashkari et~al.(2013)Lashkari, McDermott, and
  Van~Raamsdonk]{Lashkari:2013koa}
N.~Lashkari, M.~B. McDermott, and M.~Van~Raamsdonk, ``{Gravitational Dynamics
  From Entanglement ``Thermodynamics"}'',
 \href{http://xxx.lanl.gov/abs/1308.3716}{{\ttfamily arXiv:1308.3716}}.

\bibitem[Bhattacharya and Takayanagi(2013)]{Bhattacharya:2013bna}
J.~Bhattacharya and T.~Takayanagi, ``{Entropic Counterpart of Perturbative
  Einstein Equation}'',
 \href{http://xxx.lanl.gov/abs/1308.3792}{{\ttfamily arXiv:1308.3792}}.

\bibitem[Hubeny and Rangamani(2012)]{Hubeny:2012wa}
V.~E. Hubeny and M.~Rangamani, ``{Causal Holographic Information}'', {\em JHEP}
  {\bfseries 1206} (2012) 114,
 \href{http://xxx.lanl.gov/abs/1204.1698}{{\ttfamily arXiv:1204.1698}}.

\bibitem[Hubeny et~al.(2013{\natexlab{a}})Hubeny, Rangamani, and
  Tonni]{Hubeny:2013hz}
V.~E. Hubeny, M.~Rangamani, and E.~Tonni, ``{Thermalization of Causal
  Holographic Information}'', {\em JHEP} {\bfseries 1305} (2013){\natexlab{a}}
  136,
 \href{http://xxx.lanl.gov/abs/1302.0853}{{\ttfamily arXiv:1302.0853}}.

\bibitem[Hubeny et~al.(2013{\natexlab{b}})Hubeny, Rangamani, and
  Tonni]{Hubeny:2013gba}
V.~E. Hubeny, M.~Rangamani, and E.~Tonni, ``{Global properties of causal wedges
  in asymptotically AdS spacetimes}'',
 \href{http://xxx.lanl.gov/abs/1306.4324}{{\ttfamily arXiv:1306.4324}}.

\bibitem[Bousso et~al.(2012)Bousso, Leichenauer, and Rosenhaus]{Bousso:2012sj}
R.~Bousso, S.~Leichenauer, and V.~Rosenhaus, ``{Light-sheets and AdS/CFT}'',
  {\em Phys.Rev.} {\bfseries D86} (2012) 046009,
 \href{http://xxx.lanl.gov/abs/1203.6619}{{\ttfamily arXiv:1203.6619}}.

\bibitem[Czech et~al.(2012)Czech, Karczmarek, Nogueira, and
  Van~Raamsdonk]{Czech:2012bh}
B.~Czech, J.~L. Karczmarek, F.~Nogueira, and M.~Van~Raamsdonk, ``{The Gravity
  Dual of a Density Matrix}'', {\em Class.Quant.Grav.} {\bfseries 29} (2012)
  155009,
 \href{http://xxx.lanl.gov/abs/1204.1330}{{\ttfamily arXiv:1204.1330}}.

\bibitem[Hawking(1971)]{Hawking:1971tu}
S.~Hawking, ``{Gravitational radiation from colliding black holes}'', {\em
  Phys.Rev.Lett.} {\bfseries 26} (1971)
1344--1346.

\bibitem[Jacobson and Parentani(2003)]{Jacobson:2003wv}
T.~Jacobson and R.~Parentani, ``{Horizon entropy}'', {\em Found.Phys.}
  {\bfseries 33} (2003) 323--348,
 \href{http://xxx.lanl.gov/abs/gr-qc/0302099}{{\ttfamily arXiv:gr-qc/0302099}}.

\bibitem[Geroch(1970)]{Geroch:1970uw}
R.~P. Geroch, ``{The domain of dependence}'', {\em J.Math.Phys.} {\bfseries 11}
  (1970)
437--439.

\bibitem[Kabat(1995)]{Kabat:1995eq}
D.~N. Kabat, ``{Black hole entropy and entropy of entanglement}'', {\em
  Nucl.Phys.} {\bfseries B453} (1995) 281--302,
 \href{http://xxx.lanl.gov/abs/hep-th/9503016}{{\ttfamily
  arXiv:hep-th/9503016}}.

\bibitem[Larsen and Wilczek(1996)]{Larsen:1995ax}
F.~Larsen and F.~Wilczek, ``{Renormalization of black hole entropy and of the
  gravitational coupling constant}'', {\em Nucl.Phys.} {\bfseries B458} (1996)
  249--266,
 \href{http://xxx.lanl.gov/abs/hep-th/9506066}{{\ttfamily
  arXiv:hep-th/9506066}}.

\bibitem[Iellici and Moretti(1996)]{Iellici:1996jx}
D.~Iellici and V.~Moretti, ``{Kabat's surface terms in the zeta function
  approach}'',
 \href{http://xxx.lanl.gov/abs/hep-th/9703088}{{\ttfamily
  arXiv:hep-th/9703088}}.

\bibitem[Zhitnitsky(2011)]{Zhitnitsky:2011tr}
A.~R. Zhitnitsky, ``{Entropy, Contact Interaction with Horizon and Dark
  Energy}'', {\em Phys.Rev.} {\bfseries D84} (2011) 124008,
 \href{http://xxx.lanl.gov/abs/1105.6088}{{\ttfamily arXiv:1105.6088}}.

\bibitem[Donnelly and Wall(2012)]{Donnelly:2012st}
W.~Donnelly and A.~C. Wall, ``{Do gauge fields really contribute negatively to
  black hole entropy?}'', {\em Phys.Rev.} {\bfseries D86} (2012) 064042,
 \href{http://xxx.lanl.gov/abs/1206.5831}{{\ttfamily arXiv:1206.5831}}.

\bibitem[Solodukhin(2012)]{Solodukhin:2012jh}
S.~N. Solodukhin, ``{Remarks on effective action and entanglement entropy of
  Maxwell field in generic gauge}'', {\em JHEP} {\bfseries 1212} (2012) 036,
 \href{http://xxx.lanl.gov/abs/1209.2677}{{\ttfamily arXiv:1209.2677}}.

\bibitem[Eling et~al.(2013)Eling, Oz, and Theisen]{Eling:2013aqa}
C.~Eling, Y.~Oz, and S.~Theisen, ``{Entanglement and Thermal Entropy of Gauge
  Fields}'',
 \href{http://xxx.lanl.gov/abs/1308.4964}{{\ttfamily arXiv:1308.4964}}.

\bibitem[Donnelly and Wall(2013)]{Donnelly:2013}
W.~Donnelly and A.~Wall,  2013, unpublished.

\bibitem[Susskind and Witten(1998)]{Susskind:1998dq}
L.~Susskind and E.~Witten, ``{The Holographic bound in anti-de Sitter space}'',
 \href{http://xxx.lanl.gov/abs/hep-th/9805114}{{\ttfamily
  arXiv:hep-th/9805114}}.

\bibitem[Freivogel and Mosk(2013)]{Freivogel:2013zta}
B.~Freivogel and B.~Mosk, ``{Properties of Causal Holographic Information}'',
 \href{http://xxx.lanl.gov/abs/1304.7229}{{\ttfamily arXiv:1304.7229}}.

\bibitem[Gell-Mann and Hartle(2007)]{GellMann:2006uj}
M.~Gell-Mann and J.~Hartle, ``{Quasiclassical Coarse Graining and Thermodynamic
  Entropy}'', {\em Phys.Rev.} {\bfseries A76} (2007) 022104,
 \href{http://xxx.lanl.gov/abs/quant-ph/0609190}{{\ttfamily
  arXiv:quant-ph/0609190}}.

\bibitem[Bekenstein(1973)]{Bekenstein:1973ur}
J.~D. Bekenstein, ``{Black holes and entropy}'', {\em Phys.Rev.} {\bfseries D7}
  (1973)
2333--2346.

\bibitem[Hawking(1975)]{Hawking:1974sw}
S.~Hawking, ``{Particle Creation by Black Holes}'', {\em Commun.Math.Phys.}
  {\bfseries 43} (1975)
199--220.

\bibitem[Wall(2009)]{Wall:2009wm}
A.~C. Wall, ``{Ten Proofs of the Generalized Second Law}'', {\em JHEP}
  {\bfseries 0906} (2009) 021,
 \href{http://xxx.lanl.gov/abs/0901.3865}{{\ttfamily arXiv:0901.3865}}.

\bibitem[Henningson and Skenderis(1998)]{Henningson:1998gx}
M.~Henningson and K.~Skenderis, ``{The Holographic Weyl anomaly}'', {\em JHEP}
  {\bfseries 9807} (1998) 023,
 \href{http://xxx.lanl.gov/abs/hep-th/9806087}{{\ttfamily
  arXiv:hep-th/9806087}}.

\bibitem[Balasubramanian and Kraus(1999)]{Balasubramanian:1999re}
V.~Balasubramanian and P.~Kraus, ``{A Stress tensor for Anti-de Sitter
  gravity}'', {\em Commun.Math.Phys.} {\bfseries 208} (1999) 413--428,
 \href{http://xxx.lanl.gov/abs/hep-th/9902121}{{\ttfamily
  arXiv:hep-th/9902121}}.

\bibitem[Skenderis(2002)]{Skenderis:2002wp}
K.~Skenderis, ``{Lecture notes on holographic renormalization}'', {\em
  Class.Quant.Grav.} {\bfseries 19} (2002) 5849--5876,
 \href{http://xxx.lanl.gov/abs/hep-th/0209067}{{\ttfamily
  arXiv:hep-th/0209067}}.

\bibitem[{Sorkin}(1986)]{Sorkin:1986}
R.~D. {Sorkin}, ``{Introduction to Topological Geons}'', in ``NATO ASIB Proc.
  138: Topological Structure of Space-Time'', P.~G. {Bergmann} and V.~{de
  Sabbata}, eds., p.~249.
\newblock 1986.

\bibitem[Friedman et~al.(1993)Friedman, Schleich, and Witt]{Friedman:1993ty}
J.~L. Friedman, K.~Schleich, and D.~M. Witt, ``{Topological censorship}'', {\em
  Phys.Rev.Lett.} {\bfseries 71} (1993) 1486--1489,
 \href{http://xxx.lanl.gov/abs/gr-qc/9305017}{{\ttfamily arXiv:gr-qc/9305017}}.

\bibitem[Louko and Marolf(1999)]{Louko:1998hc}
J.~Louko and D.~Marolf, ``{Single exterior black holes and the AdS / CFT
  conjecture}'', {\em Phys.Rev.} {\bfseries D59} (1999) 066002,
 \href{http://xxx.lanl.gov/abs/hep-th/9808081}{{\ttfamily
  arXiv:hep-th/9808081}}.

\bibitem[Smith and Mann(2013)]{Smith:2013zqa}
A.~R.~H. Smith and R.~B. Mann, ``{Looking Inside a Black Hole}'',
 \href{http://xxx.lanl.gov/abs/1309.4125}{{\ttfamily arXiv:1309.4125}}.

\bibitem[Gao and Wald(2000)]{Gao:2000ga}
S.~Gao and R.~M. Wald, ``{Theorems on gravitational time delay and related
  issues}'', {\em Class.Quant.Grav.} {\bfseries 17} (2000) 4999--5008,
 \href{http://xxx.lanl.gov/abs/gr-qc/0007021}{{\ttfamily arXiv:gr-qc/0007021}}.

\bibitem[Balasubramanian et~al.(1999{\natexlab{a}})Balasubramanian, Kraus, and
  Lawrence]{Balasubramanian:1998sn}
V.~Balasubramanian, P.~Kraus, and A.~E. Lawrence, ``{Bulk versus boundary
  dynamics in anti-de Sitter space-time}'', {\em Phys.Rev.} {\bfseries D59}
  (1999){\natexlab{a}} 046003,
 \href{http://xxx.lanl.gov/abs/hep-th/9805171}{{\ttfamily
  arXiv:hep-th/9805171}}.

\bibitem[Balasubramanian et~al.(1999{\natexlab{b}})Balasubramanian, Kraus,
  Lawrence, and Trivedi]{Balasubramanian:1998de}
V.~Balasubramanian, P.~Kraus, A.~E. Lawrence, and S.~P. Trivedi, ``{Holographic
  probes of anti-de Sitter space-times}'', {\em Phys.Rev.} {\bfseries D59}
  (1999){\natexlab{b}} 104021,
 \href{http://xxx.lanl.gov/abs/hep-th/9808017}{{\ttfamily
  arXiv:hep-th/9808017}}.

\bibitem[Banks et~al.(1998)Banks, Douglas, Horowitz, and
  Martinec]{Banks:1998dd}
T.~Banks, M.~R. Douglas, G.~T. Horowitz, and E.~J. Martinec, ``{AdS dynamics
  from conformal field theory}'',
 \href{http://xxx.lanl.gov/abs/hep-th/9808016}{{\ttfamily
  arXiv:hep-th/9808016}}.

\bibitem[Bena(2000)]{Bena:1999jv}
I.~Bena, ``{On the construction of local fields in the bulk of AdS(5) and other
  spaces}'', {\em Phys.Rev.} {\bfseries D62} (2000) 066007,
 \href{http://xxx.lanl.gov/abs/hep-th/9905186}{{\ttfamily
  arXiv:hep-th/9905186}}.

\bibitem[Hamilton et~al.(2006{\natexlab{a}})Hamilton, Kabat, Lifschytz, and
  Lowe]{Hamilton:2005ju}
A.~Hamilton, D.~N. Kabat, G.~Lifschytz, and D.~A. Lowe, ``{Local bulk operators
  in AdS/CFT: A Boundary view of horizons and locality}'', {\em Phys.Rev.}
  {\bfseries D73} (2006){\natexlab{a}} 086003,
 \href{http://xxx.lanl.gov/abs/hep-th/0506118}{{\ttfamily
  arXiv:hep-th/0506118}}.

\bibitem[Hamilton et~al.(2006{\natexlab{b}})Hamilton, Kabat, Lifschytz, and
  Lowe]{Hamilton:2006az}
A.~Hamilton, D.~N. Kabat, G.~Lifschytz, and D.~A. Lowe, ``{Holographic
  representation of local bulk operators}'', {\em Phys.Rev.} {\bfseries D74}
  (2006){\natexlab{b}} 066009,
 \href{http://xxx.lanl.gov/abs/hep-th/0606141}{{\ttfamily
  arXiv:hep-th/0606141}}.

\bibitem[Kabat et~al.(2011)Kabat, Lifschytz, and Lowe]{Kabat:2011rz}
D.~Kabat, G.~Lifschytz, and D.~A. Lowe, ``{Constructing local bulk observables
  in interacting AdS/CFT}'', {\em Phys.Rev.} {\bfseries D83} (2011) 106009,
 \href{http://xxx.lanl.gov/abs/1102.2910}{{\ttfamily arXiv:1102.2910}}.

\bibitem[Heemskerk et~al.(2012)Heemskerk, Marolf, Polchinski, and
  Sully]{Heemskerk:2012mn}
I.~Heemskerk, D.~Marolf, J.~Polchinski, and J.~Sully, ``{Bulk and Transhorizon
  Measurements in AdS/CFT}'', {\em JHEP} {\bfseries 1210} (2012) 165,
 \href{http://xxx.lanl.gov/abs/1201.3664}{{\ttfamily arXiv:1201.3664}}.

\bibitem[Bizon and Rostworowski(2011)]{Bizon:2011gg}
P.~Bizon and A.~Rostworowski, ``{On weakly turbulent instability of anti-de
  Sitter space}'', {\em Phys.Rev.Lett.} {\bfseries 107} (2011) 031102,
 \href{http://xxx.lanl.gov/abs/1104.3702}{{\ttfamily arXiv:1104.3702}}.

\bibitem[Jalmuzna et~al.(2011)Jalmuzna, Rostworowski, and
  Bizon]{Jalmuzna:2011qw}
J.~Jalmuzna, A.~Rostworowski, and P.~Bizon, ``{A Comment on AdS collapse of a
  scalar field in higher dimensions}'', {\em Phys.Rev.} {\bfseries D84} (2011)
  085021,
 \href{http://xxx.lanl.gov/abs/1108.4539}{{\ttfamily arXiv:1108.4539}}.

\bibitem[Dias et~al.(2012)Dias, Horowitz, and Santos]{Dias:2011ss}
O.~J. Dias, G.~T. Horowitz, and J.~E. Santos, ``{Gravitational Turbulent
  Instability of Anti-de Sitter Space}'', {\em Class.Quant.Grav.} {\bfseries
  29} (2012) 194002,
 \href{http://xxx.lanl.gov/abs/1109.1825}{{\ttfamily arXiv:1109.1825}}.

\bibitem[Balasubramanian et~al.(2013)Balasubramanian, Chowdhury, Czech,
  de~Boer, and Heller]{Balasubramanian:2013lsa}
V.~Balasubramanian, B.~D. Chowdhury, B.~Czech, J.~de~Boer, and M.~P. Heller,
  ``{A hole-ographic spacetime}'',
 \href{http://xxx.lanl.gov/abs/1310.4204}{{\ttfamily arXiv:1310.4204}}.

\bibitem[Maldacena(1998)]{Maldacena:1998im}
J.~M. Maldacena, ``{Wilson loops in large N field theories}'', {\em
  Phys.Rev.Lett.} {\bfseries 80} (1998) 4859--4862,
 \href{http://xxx.lanl.gov/abs/hep-th/9803002}{{\ttfamily
  arXiv:hep-th/9803002}}.

\bibitem[Rey and Yee(2001)]{Rey:1998ik}
S.-J. Rey and J.-T. Yee, ``{Macroscopic strings as heavy quarks in large N
  gauge theory and anti-de Sitter supergravity}'', {\em Eur.Phys.J.} {\bfseries
  C22} (2001) 379--394,
 \href{http://xxx.lanl.gov/abs/hep-th/9803001}{{\ttfamily
  arXiv:hep-th/9803001}}.

\bibitem[Hubeny(2012)]{Hubeny:2012ry}
V.~E. Hubeny, ``{Extremal surfaces as bulk probes in AdS/CFT}'', {\em JHEP}
  {\bfseries 1207} (2012) 093,
 \href{http://xxx.lanl.gov/abs/1203.1044}{{\ttfamily arXiv:1203.1044}}.

\bibitem[Louko et~al.(2000)Louko, Marolf, and Ross]{Louko:2000tp}
J.~Louko, D.~Marolf, and S.~F. Ross, ``{On geodesic propagators and black hole
  holography}'', {\em Phys.Rev.} {\bfseries D62} (2000) 044041,
 \href{http://xxx.lanl.gov/abs/hep-th/0002111}{{\ttfamily
  arXiv:hep-th/0002111}}.

\bibitem[Giddings and Lippert(2002)]{Giddings:2001pt}
S.~B. Giddings and M.~Lippert, ``{Precursors, black holes, and a locality
  bound}'', {\em Phys.Rev.} {\bfseries D65} (2002) 024006,
 \href{http://xxx.lanl.gov/abs/hep-th/0103231}{{\ttfamily
  arXiv:hep-th/0103231}}.

\bibitem[Gibbons and Warner(2013)]{Gibbons:2013tqa}
G.~Gibbons and N.~Warner, ``{Global Structure of Five-dimensional BPS
  Fuzzballs}'',
 \href{http://xxx.lanl.gov/abs/1305.0957}{{\ttfamily arXiv:1305.0957}}.

\bibitem[Giddings and Ross(2000)]{Giddings:1999zu}
S.~B. Giddings and S.~F. Ross, ``{D3-brane shells to black branes on the
  Coulomb branch}'', {\em Phys.Rev.} {\bfseries D61} (2000) 024036,
 \href{http://xxx.lanl.gov/abs/hep-th/9907204}{{\ttfamily
  arXiv:hep-th/9907204}}.

\bibitem[Chepelev and Roiban(1999)]{Chepelev:1999zt}
I.~Chepelev and R.~Roiban, ``{A Note on correlation functions in AdS(5) /
  SYM(4) correspondence on the Coulomb branch}'', {\em Phys.Lett.} {\bfseries
  B462} (1999) 74--80,
 \href{http://xxx.lanl.gov/abs/hep-th/9906224}{{\ttfamily
  arXiv:hep-th/9906224}}.

\bibitem[Bousso et~al.(2012)Bousso, Freivogel, Leichenauer, Rosenhaus, and
  Zukowski]{Bousso:2012mh}
R.~Bousso, B.~Freivogel, S.~Leichenauer, V.~Rosenhaus, and C.~Zukowski, ``{Null
  Geodesics, Local CFT Operators and AdS/CFT for Subregions}'',
 \href{http://xxx.lanl.gov/abs/1209.4641}{{\ttfamily arXiv:1209.4641}}.

\bibitem[Leichenauer and Rosenhaus(2013)]{Leichenauer:2013kaa}
S.~Leichenauer and V.~Rosenhaus, ``{AdS black holes, the bulk-boundary
  dictionary, and smearing functions}'', {\em Phys.Rev.} {\bfseries D88} (2013)
  026003,
 \href{http://xxx.lanl.gov/abs/1304.6821}{{\ttfamily arXiv:1304.6821}}.

\bibitem[Czech et~al.(2012)Czech, Karczmarek, Nogueira, and
  Van~Raamsdonk]{Czech:2012be}
B.~Czech, J.~L. Karczmarek, F.~Nogueira, and M.~Van~Raamsdonk, ``{Rindler
  Quantum Gravity}'', {\em Class.Quant.Grav.} {\bfseries 29} (2012) 235025,
 \href{http://xxx.lanl.gov/abs/1206.1323}{{\ttfamily arXiv:1206.1323}}.

\bibitem[Hubeny et~al.(2013)Hubeny, Maxfield, Rangamani, and
  Tonni]{Hubeny:2013gta}
V.~E. Hubeny, H.~Maxfield, M.~Rangamani, and E.~Tonni, ``{Holographic
  entanglement plateaux}'',
 \href{http://xxx.lanl.gov/abs/1306.4004}{{\ttfamily arXiv:1306.4004}}.

\bibitem[Hubeny et~al.(2010{\natexlab{a}})Hubeny, Marolf, and
  Rangamani]{Hubeny:2009ru}
V.~E. Hubeny, D.~Marolf, and M.~Rangamani, ``{Hawking radiation in large N
  strongly-coupled field theories}'', {\em Class.Quant.Grav.} {\bfseries 27}
  (2010){\natexlab{a}} 095015,
 \href{http://xxx.lanl.gov/abs/0908.2270}{{\ttfamily arXiv:0908.2270}}.

\bibitem[Hubeny et~al.(2010{\natexlab{b}})Hubeny, Marolf, and
  Rangamani]{Hubeny:2009kz}
V.~E. Hubeny, D.~Marolf, and M.~Rangamani, ``{Black funnels and droplets from
  the AdS C-metrics}'', {\em Class.Quant.Grav.} {\bfseries 27}
  (2010){\natexlab{b}} 025001,
 \href{http://xxx.lanl.gov/abs/0909.0005}{{\ttfamily arXiv:0909.0005}}.

\bibitem[Hubeny et~al.(2010{\natexlab{c}})Hubeny, Marolf, and
  Rangamani]{Hubeny:2009rc}
V.~E. Hubeny, D.~Marolf, and M.~Rangamani, ``{Hawking radiation from AdS black
  holes}'', {\em Class.Quant.Grav.} {\bfseries 27} (2010){\natexlab{c}} 095018,
 \href{http://xxx.lanl.gov/abs/0911.4144}{{\ttfamily arXiv:0911.4144}}.

\bibitem[Santos and Way(2012)]{Santos:2012he}
J.~E. Santos and B.~Way, ``{Black Funnels}'', {\em JHEP} {\bfseries 1212}
  (2012) 060,
 \href{http://xxx.lanl.gov/abs/1208.6291}{{\ttfamily arXiv:1208.6291}}.

\bibitem[Wall(2013)]{Wall:2010st}
A.~C. Wall, ``{The Generalized Second Law implies a Quantum Singularity
  Theorem}'', {\em Class.Quant.Grav.} {\bfseries 30} (2013) 165003,
 \href{http://xxx.lanl.gov/abs/1010.5513}{{\ttfamily arXiv:1010.5513}}.

\bibitem[Dias et~al.(2012)Dias, Horowitz, Marolf, and Santos]{Dias:2012tq}
O.~J. Dias, G.~T. Horowitz, D.~Marolf, and J.~E. Santos, ``{On the Nonlinear
  Stability of Asymptotically Anti-de Sitter Solutions}'', {\em
  Class.Quant.Grav.} {\bfseries 29} (2012) 235019,
 \href{http://xxx.lanl.gov/abs/1208.5772}{{\ttfamily arXiv:1208.5772}}.

\bibitem[Marolf(2009)]{Marolf:2008mg}
D.~Marolf, ``{Holographic Thought Experiments}'', {\em Phys.Rev.} {\bfseries
  D79} (2009) 024029,
 \href{http://xxx.lanl.gov/abs/0808.2845}{{\ttfamily arXiv:0808.2845}}.

\bibitem[Balasubramanian et~al.(2013)Balasubramanian, Czech, Chowdhury, and
  de~Boer]{Balasubramanian:2013rqa}
V.~Balasubramanian, B.~Czech, B.~D. Chowdhury, and J.~de~Boer, ``{The entropy
  of a hole in spacetime}'',
 \href{http://xxx.lanl.gov/abs/1305.0856}{{\ttfamily arXiv:1305.0856}}.

\bibitem[Faulkner et~al.(2013)Faulkner, Lewkowycz, and
  Maldacena]{Faulkner:2013ana}
T.~Faulkner, A.~Lewkowycz, and J.~Maldacena, ``{Quantum corrections to
  holographic entanglement entropy}'',
 \href{http://xxx.lanl.gov/abs/1307.2892}{{\ttfamily arXiv:1307.2892}}.

\bibitem[Wall(2012)]{Wall:2011hj}
A.~C. Wall, ``{A proof of the generalized second law for rapidly changing
  fields and arbitrary horizon slices}'', {\em Phys.Rev.} {\bfseries D85}
  (2012) 104049,
 \href{http://xxx.lanl.gov/abs/1105.3445}{{\ttfamily arXiv:1105.3445}}.

\end{thebibliography}\endgroup

\end{document}